\documentclass{article}

\usepackage[preprint,nonatbib]{neurips_data_2022}

\usepackage[utf8]{inputenc} 
\usepackage[T1]{fontenc}    
\usepackage{hyperref}       
\usepackage{url}            
\usepackage{booktabs}       
\usepackage{amsfonts}       
\usepackage{nicefrac}       
\usepackage{microtype}      
\usepackage{xcolor}         
\usepackage{tabu,multirow}

\newcommand{\systemnameAPIBenchmark}{\textsc{HAPI}}

\title{\systemnameAPIBenchmark: A Large-scale Longitudinal Dataset of Commercial ML API Predictions}

\author{%
Lingjiao Chen$^1$, Zhihua Jin$^2$,  Sabri Eyuboglu$^1$, Christopher R\'e$^1$,  Matei Zaharia$^1$, James Zou$^1$
 \\ Stanford University$^1$, Hong Kong University of Science and Technology$^2$ \\
}

\usepackage{wrapfig}
\usepackage{bbm}
\usepackage{hyperref}       
\usepackage{url}            
\usepackage{booktabs}       
\usepackage{amsfonts}       
\usepackage{microtype}      
\usepackage{microtype}
\usepackage{graphicx}
\usepackage{subfigure}
\usepackage{booktabs} 
\usepackage{amsthm,amsmath,amssymb}
\usepackage{float,url,amsfonts,alltt}
\usepackage{mathtools,rotating}
\usepackage{ifpdf,fancyvrb}
\usepackage{enumitem}

\usepackage{microtype}
\usepackage{graphicx}
\usepackage{booktabs} 
\usepackage{hyperref}

\usepackage{graphicx,xspace,verbatim,comment}
\usepackage{hyperref,array,color,balance,multirow}
\usepackage{balance,float,url,amsfonts,alltt}
\usepackage{mathtools,rotating,amsmath,amssymb}
\usepackage{color,ifpdf,fancyvrb,array}
\usepackage{etoolbox,listings}
\usepackage{bigstrut,morefloats}

\usepackage{pbox}
\usepackage[boxruled,algo2e,linesnumbered]{algorithm2e}
\DeclarePairedDelimiterX{\inp}[2]{\langle}{\rangle}{#1, #2}

\newcommand{\eat}[1]{}

\numberwithin{equation}{section}

\usepackage{amsmath}
\usepackage{accents}
\newlength{\dhatheight}

\usepackage{algorithm,algpseudocode}

\usepackage[T1]{fontenc}
\usepackage{upgreek}

\usepackage[greek,english]{babel}

\newcommand{\PO}{ \textit{PO} } 
\newcommand{\ConfMove}{ \textit{CM} } 
\newcommand{\Acc}{ a } 
\newcommand{\GD}{ \textit{GD} } 

\newcommand{\MultiAcc}{ ma }

\begin{document}

\maketitle

\begin{abstract}
Commercial ML APIs offered by providers such as Google, Amazon and Microsoft have dramatically simplified ML adoption in many applications. Numerous companies and academics pay to use ML APIs for tasks such as object detection, OCR and sentiment analysis. Different ML APIs tackling the same task can have very heterogeneous performance. Moreover, the ML models underlying the APIs also evolve over time. As ML APIs rapidly become a valuable marketplace and a widespread way to consume machine learning, it is critical to systematically study and compare different APIs with each other and to characterize how APIs change over time. However, this topic is currently underexplored due to the lack of data. In this paper, we present \systemnameAPIBenchmark{} (History of APIs), a longitudinal dataset of        
1,761,417 instances of commercial ML API applications (involving APIs from Amazon, Google, IBM,  Microsoft and other providers) across diverse tasks including image tagging, speech recognition and text mining from 2020 to 2022. Each instance consists of a query input for an API (e.g., an image or text) along with the API's output prediction/annotation and confidence scores. \systemnameAPIBenchmark{} is the first large-scale dataset of ML API usages and is a unique resource for studying ML-as-a-service (MLaaS). As examples of the types of analyses that \systemnameAPIBenchmark{} enables, we show that ML APIs' performance change substantially over time---several APIs' accuracies dropped on specific benchmark datasets. Even when the API's aggregate performance stays steady, its error modes can shift across different subtypes of data between 2020 and 2022. Such changes can substantially impact the entire analytics pipelines that use some ML API as a component. We further use \systemnameAPIBenchmark{} to study  commercial APIs' performance disparities across demographic subgroups over time.   
\systemnameAPIBenchmark{} can stimulate more research in the growing field of MLaaS.
\end{abstract}

\section{Introduction}
\label{sec:introduction}
Machine learning (ML) prediction APIs have dramatically simplified ML adoption.
For example, one can use the Google speech API to transform an utterance to a text paragraph, or the Microsoft vision API to recognize all objects in an image. The ML-as-a-Service (MLaaS) market powered by these APIs is increasingly growing and expected to exceed \$16 billion USD in the next five years~\cite{MLasS_MarketInfo}.

Despite its increasing popularity, systematic analysis of this MLaaS ecosystem is limited, and many phenomena are not well understood. 
For example, APIs from different providers can have heterogeneous performance on the same dataset. Deciding which API or combination of APIs to use on a specific dataset can be challenging. Moreover, providers can update their ML APIs due to new data availability and model architecture advancements, but users often do not know how the API's behavior on their data changes. 
Such API shifts can substantially affect (and hurt) the performance of downstream applications.  
Certain biases or stereotypes in the ML APIs~\cite{pmlr-v81-buolamwini18a} can also be amplified or mitigated by API shifts. 
Understanding the dynamics of ML APIs is critical for ensuring the reliability of the entire user pipeline, for which the API is one component. It also helps users to adjust their API usage strategies timely and appropriately. 
For example, one may trust a speech API's prediction if its confidence score is higher than 90\% and invoke a human expert otherwise. 
Suppose the API is updated so that its confidence is reduced by 10\% while its prediction remains the same (this happens in practice, as we will show).
Then the human invocation threshold also needs to be adjusted to ensure consistent overall performance and human workload.

\textbf{Our contributions} In this paper, we present \systemnameAPIBenchmark{} (History of APIs), a longitudinal dataset of 1,761,417
 data points annotated by a range of different ML APIs from Google, Microsoft, Amazon and other providers from 2020 to 2022.
This covers ML APIs for both standard classification such as sentiment analysis and structured prediction tasks including multi-label image classification.
We have released our dataset on the project website~\footnote{\url{https://github.com/lchen001/HAPI/}}, and will keep updating it by querying all ML APIs every few months.
To the best of our knowledge, \systemnameAPIBenchmark{} is the first systematic dataset of ML API applications. It is a unique resource that facilitates studies of the increasingly critical MLaaS ecosystem. Furthermore, we use \systemnameAPIBenchmark{} 
to characterize interesting findings on API shifts between 2020 and 2022. 
Our analysis shows that API shifts are common: more than 60\% of the 63 evaluated API-dataset pairs encounter performance shifts. 
Those API shifts lead to both accuracy improvements and drops. 
For example, Google vision API's shift from 2020 to 2022 brings a 1\% performance drop on the PASCAL dataset but a 3.7\% improvement on the MIR data.  
Interestingly, the fraction of changed predictions is often larger than the accuracy change, indicating that 
an API update may fix certain mistakes but introduce additional errors.
ML APIs' confidence scores can also change even if the predictions do not.
For example, from 2020 to 2021, the average confidence score of the Microsoft speech API increased by 30\%  while its accuracy became lower; in contrast, IBM API's confidence dropped by 1\% but its accuracy actually improved. 
We also observe that subgroup performance disparity produced by different ML APIs is consistent over time. \systemnameAPIBenchmark{} provides a rich resource to stimulate research on the under-explored but increasingly important topic of MLaaS. 

\section{Related Work}
\label{sec:related_work}
To the best of our knowledge, \systemnameAPIBenchmark{} is the first large-scale ML API dataset. We discuss relevant literature below. 

\paragraph{MLaaS.}

MLaaS APIs \cite{FrugalML2020} have been developed and sold by giant companies including Google~\cite{GoogleAPI} and  Amazon~\cite{AmazonAPI} as well as startups such as Face++~\cite{FacePPAPI} and EPixel~\cite{EverypixelAPI}.
Many applications have been discovered~\cite{ pmlr-v81-buolamwini18a, MLasS_GoogleDigitalMedia_2019, MLasS_Google_Microsoft_Blind18}, and prior work on ML APIs has spanned on their robustness~\cite{MLasS_GoogleNotRobust_2017} and pricing mechanisms~\cite{nimbus_chen2019towards}. One challenge in MLaaS is  to determine which API or combination of them to use given a user budget constraint.  
This requires adaptive API calling strategies to jointly consider performance and cost, studied in recent work such as FrugalML~\cite{FrugalML2020} and FrugalMCT~\cite{chen2021frugalmct}.
While they also released datasets of ML API predictions, their dataset only contain evaluation in one year. 
Recent work on API shift estimation~\cite{chen2022APIShift} evaluated a few classification ML APIs in two years. 
On the other hand, \systemnameAPIBenchmark{} provides a systematical evaluation of a number of ML APIs over a couple of years and thus enables more research on ML APIs evolution over time. 
\paragraph{Dynamics of ML systems.} ML is a fast growing community~\cite{schmidhuber2015deep} and the update of one component may impact an ML system significantly. 
For example, a recent study on dataset dynamics~\cite{koch2021reduced} implies a concentration on fewer and fewer datasets over time and thus potentially increasing biases in many ML systems.
Various hardware  optimization~\cite{sze2017hardware} are shown to accelerate training and inference speeds for many applications.  
\systemnameAPIBenchmark{} focuses on the dynamics of ML APIs, another important component of many ML systems.

\paragraph{ML pipeline monitoring and assessments.}
Monitoring and assessing ML pipelines are critical in real world ML applications. 
Existing work studies on how to estimate the performance of a deployed ML model based on certain statistics such as confidence~\cite{modeleval_confidence_guillory2021predicting}, rotation prediction~\cite{ModelEval_rotation_deng2021does} and feature statistics of the datasets sampled from a meta-dataset~\cite{ModelEval_featurestats_deng2021labels}.
More general approaches exploit human knowledge~\cite{ModelEval_mandoline_chen2021}, white-box access to the ML models~\cite{ModelEval_chen2021detecting}, or known label or feature distribution~\cite{ModelEval_donmez2010unsupervised, ModelEval_MIT_chuang2020estimating}. 
Another line of work is identifying errors made by an ML model.
This involves ML models for tabular data~\cite{breck2019data,baylor2017tfx} as well as multimedia data~\cite{kang2020model}.
One common assumption made by them is that the deployed ML models are fixed and the performance change or error emergence is due to data distribution shifts. 
However, our analysis on \systemnameAPIBenchmark{} indicates that ML systems powered by ML APIs may also change notably.
This calls for monitoring and assessments under both model and data distribution shifts.

\eat{
The rest of the paper is organized as follows. 
Section \ref{} gives the preliminary and related work. 
We describe how \systemnameAPIBenchmark{} is constructed  in Section \ref{} and present a detailed analysis on \systemnameAPIBenchmark{} in Section \ref{}. 
Section \ref{} offers open questions and challenges enabled by \systemnameAPIBenchmark{}, and finally we conclude the paper in Section \ref{}. 
}
\section{Construction of \systemnameAPIBenchmark{}: Tasks, Datasets, and ML APIs}
\label{sec:construction}

\begin{table}[htbp]
  \centering
  \small
  \caption{Evaluated ML APIs. For each task, we have evaluated three popular  ML APIs from different commercial providers. The valuation was conducted in the spring of 2020, 2021, and 2022 for classification tasks, and 2020 fall as well as 2022 spring for structured prediction tasks.}
    \begin{tabular}{|c||c|c|c|c|c|}
    \hline
    Task Type & Task  & \multicolumn{3}{c|}{ML API} & \multicolumn{1}{c|}{Evaluation Period} \bigstrut\\
    \hline
    \hline
    \multirow{3}[6]{*}{Classify} & SCR   & Google~\cite{GoogleSpeechAPI} & Microsoft~\cite{MicrosoftSpeechAPI} & IBM~\cite{IBMAPI}   & March 2020, April 2021, May 2022 \bigstrut\\
\cline{2-6}          & SA    & Google~\cite{GoogleNLPAPI} & Amazon~\cite{AmazonAPI} & Baidu~\cite{BaiduAPI} & March 2020, Feb 2021, May 2022 \bigstrut\\
\cline{2-6}          & FER   & Google~\cite{GoogleAPI} & Microsoft~\cite{MicrosoftAPI} & Face++~\cite{FacePPAPI} & March 2020, Feb 2021, May 2022 \bigstrut\\
    \hline
    \multirow{3}[6]{*}{Struc Pred} & MIC   & Google~\cite{GoogleAPI} & Microsoft~\cite{MicrosoftAPI} & EPixel~\cite{EverypixelAPI} & October 2020, Feb 2022 \bigstrut\\
\cline{2-6}          & STR   & Google~\cite{GoogleAPI} & iFLYTEK~\cite{IflytekAPI} & Tencent~\cite{TencentAPI}   & September 2020, March 2022 \bigstrut\\
\cline{2-6}          & NER   & Google~\cite{GoogleNLPAPI} & Amazon~\cite{AmazonAPI} & IBM~\cite{IBMNLPAPI}   & September 2020, March 2022 \bigstrut\\
    \hline
    \end{tabular}%
  \label{tab:APIBenchmark:MLAPIs}%
\end{table}%

\begin{table}[htbp]
  \centering
  \small
  \caption{Prices of ML services used for each task at their evaluation times. Price unit: USD/10,000 queries. We documented the price in 2020, 2021, and 2022 for standard classification tasks and 2020 and 2022 for structured predictions.  Note that for the same task, the prices of different ML APIs are diverse. 
  On the other hand, for a fixed ML API, its price is often stable over the past few years. }
    \begin{tabular}{|c|>{\centering\arraybackslash}p{0.9cm}||>{\centering\arraybackslash}p{0.5cm}|>{\centering\arraybackslash}p{0.5cm}|>{\centering\arraybackslash}p{0.5cm}||>{\centering\arraybackslash}p{1.1cm}||>{\centering\arraybackslash}p{0.5cm}|>{\centering\arraybackslash}p{0.5cm}|>{\centering\arraybackslash}p{0.5cm}||>{\centering\arraybackslash}p{0.9cm}||>{\centering\arraybackslash}p{0.5cm}|>{\centering\arraybackslash}p{0.5cm}|>{\centering\arraybackslash}p{0.5cm}|}
    \hline
    \multirow{2}[4]{*}{Task} & \multirow{2}[4]{*}{ML API} & \multicolumn{3}{c||}{Price} & \multirow{2}[4]{*}{ML API} & \multicolumn{3}{c||}{Price} & \multirow{2}[4]{*}{ML API} & \multicolumn{3}{c|}{Price} \bigstrut\\
\cline{3-5}\cline{7-9}\cline{11-13}          &       & 2020  & 2021  & 2022  &       & 2020  & 2021  & 2022  &       & 2020  & 2021  & 2022 \bigstrut\\
    \hline
    \hline
    SCR   & Google & 60    & 60    & 60    & MS & 41    & 41    & 41    & IBM   & 25    & 25    & 25 \bigstrut\\
    \hline
    SA    & Google & 2.5   & 2.5   & 2.5   & Amazon & 0.75  & 0.75  & 0.75  & Baidu & 3.5   & 3.6   & 3.7 \bigstrut\\
    \hline
    FER   & Google & 15    & 15    & 15'   & MS & 10    & 10    & 10    & Face++ & 5     & 5     & 5 \bigstrut\\
    \hline
    MIC   & Google & 15    &       & 15    & MS & 10    &       & 10    & EPixel & 6     &       & 6 \bigstrut\\
    \hline
    STR   & Google & 15    &       & 15    & iFLYTEK & 50    &       & 52    & Tencent & 210   &       & 210 \bigstrut\\
    \hline
    NER   & Google & 10    &       & 10    & Amazon & 3     &       & 3     & IBM   & 30    &       & 30 \bigstrut\\
    \hline
    \end{tabular}%
  \label{tab:APIBenchmark:APIPrice}%
\end{table}%

Let us first introduce \systemnameAPIBenchmark{}, a longitudinal dataset for ML prediction APIs. To assess ML APIs comprehensively, we designed 
\systemnameAPIBenchmark{} to include evaluations of (i) a large set of popular commercial ML APIs for (ii) diverse tasks (iii) on a range of standard benchmark datasets (iv) across multiple years.   
For (ii), we consider six different tasks in two categories: standard classification tasks including spoken command recognition (SCR), sentiment analysis (SA), and facial emotion recognition (FER), and structured predictions including multi-label image classification (MIC), scene text recognition (STR), and named entity recognition (NER). To achieve (i) and (iv), we have evaluated three different APIs from leading companies for each task from 2020 to 2022, summarized in Table \ref{tab:APIBenchmark:MLAPIs}. 
Specifically, we have evaluated all classification APIs in the spring of 2020, 2021, and 2022, separately, and  all structured prediction APIs in 2020 fall and 2022 spring respectively. 
The prices of all evaluated ML APIs are presented in Table \ref{tab:APIBenchmark:APIPrice}.
Note that, for any fixed task, the prices of different ML APIs vary in a large range. 
This implies selection of different ML APIs may impact the dollar cost of a downstream application. 
Interestingly, for a fixed ML API, there is almost no change in its price over the past few years.
We will also continuously evaluate those APIs and update \systemnameAPIBenchmark{} in the future.

\begin{table}[t]
  \centering
  \small
  \caption{Datasets used to evaluate classification APIs (in tasks SCR, SA, FER) and structured prediction APIs (in tasks MIC, STR, NER). We queried each dataset on all three APIs that are relevant for that task. }
    \begin{tabu}{|c||c|c|c|c|c|c|}
    \hline
    Task  & Dataset & Size  & \# Labels & Dataset & Size  & \# Labels \bigstrut\\
    \hline
\tabucline[1pt]{1-7}
    \hline
    \multirow{2}[4]{*}{Speech Command Recog} & DIGIT~\cite{Dataset_Speech_DIGIT} & 2000  & 10    & AMNIST~\cite{Dataset_Speech_AudioMNIST_becker2018interpreting} & 30000 & 10 \bigstrut\\
\cline{2-7}          & CMD~\cite{Dataset_Speech_GoogleCommand}   & 64727 & 31    & FLUENT~\cite{Dataset_Speech_Fluent_LugoschRITB19} & 30043 & 31 \bigstrut\\
\tabucline[1pt]{1-7}
    \hline
    \multirow{2}[4]{*}{Sentiment Analysis} & IMDB~\cite{Dataset_SEntiment_IMDB_ACL_HLT2011}  & 25000 & 2     & YELP~\cite{Dataset_SEntiment_YELP} & 20000    & 2 \bigstrut\\
\cline{2-7}          & WAIMAI~\cite{Dataset_SENTIMENT_WAIMAI} & 11987 & 2     & SHOP~\cite{Dataset_SENTIMENT_SHOP}  & 62774 & 2 \bigstrut\\
\tabucline[1pt]{1-7}
    \hline
    \multirow{2}[4]{*}{Facial Emotion Recog} & FER+~\cite{dataset_FERP_BarsoumZCZ16}  & 6358  & 7     & RAFDB~\cite{Dataset_FAFDB_li2017reliable} & 15339 & 7 \bigstrut\\
\cline{2-7}          & EXPW~\cite{Dataset_EXPW_SOCIALRELATION_ICCV2015}  & 31510 & 7     & AFNET~\cite{Dataset_AFFECTNET_MollahosseiniHM19}  & 287401 & 7 \bigstrut\\
\tabucline[1pt]{1-7}
    \hline
    \multirow{2}[4]{*}{Multi-label Image Class} & PASCAL~\cite{Dataset_Pascal_2015} & 11540 & 20    & MIR~\cite{Dataset_MIR_2008}   & 25000 & 25  \bigstrut\\
\cline{2-7}          & COCO~\cite{Dataset_COCO_2014}  & 123287 & 80    &    &   &  \bigstrut\\
\tabucline[1pt]{1-7}
    \hline
    \multirow{2}[4]{*}{Scene Text Recog} & MTWI~\cite{Dataset_MTWI_2018}  & 9742  & 4404  & ReCTS~\cite{Dataset_ReCTS_2019} & 20000 & 4134   \bigstrut\\
\cline{2-7}          & LSVT~\cite{Dataset_LSVT_2019}  & 30000 & 4852    &   &  &  \bigstrut\\
\tabucline[1pt]{1-7}
    \hline
    \multirow{2}[4]{*}{Named Entity Recog} & CONLL~\cite{Dataset_CONLL_2003} & 10898 & 9910  & GMB~\cite{Dataset_GMB_2013}   & 47830 & 14376  \bigstrut\\
\cline{2-7}          & ZHNER~\cite{ZHNER_dataset_github} & 16915 & 4375     &   &  &  \bigstrut\\
\tabucline[1pt]{1-7}
    \hline
    \end{tabu}%
  \label{tab:APIBenchmark:DatasetClassification}%
\end{table}%

What remains is on which datasets the ML APIs have been evaluated. To ensure (iii),    we choose four commonly-used benchmark datasets for each classification task, and three datasets for each structure prediction task. 
The dataset statistics are summarized in Table \ref{tab:APIBenchmark:DatasetClassification}. 
Note that those datasets are diverse in their size and number of labels, and thus we hope they can represent a large range of real world ML API use cases.
Some datasets come with additional meta data. For example, the speaker accents are available for the spoken command dataset DIGIT. 
Such information can be used to study how an ML API's bias changes over time. We leave more details in the appendix.

The output formats of different ML APIs are often different. 
For example, Google API generates a Google client object for each input data while Everypixel API simply returns a dictionary. To mitigate such heterogeneity, we propose a simple abstraction to represent an ML API's output.
given each data point $x$ and 
evaluation time $t$, a classification ML API's output is  (i) a predicted label $f(x,t)$  and (ii) the associated confidence score $q(x,t)$.
For structured prediction tasks, the output includes (i) a set of predicted labels $f(x,t)$ (ii) associated with their quality scores $q(x,t)$. For each ML API and dataset pair, we recorded the API's prediction $f(x,t)$ and $q(x,t)$ at each evaluation time. We also include the true label $y$ for each dataset. 

As a result, \systemnameAPIBenchmark{} consists of 1,761,417
 data samples from various tasks and datasets annotated by commercial ML APIs from 2020 to 2022.
 We provide download access on the project website, and also offer a few interesting examples for exploration purposes. 

\section{Example Analyses Enabled by \systemnameAPIBenchmark{}: Model Shifts Over Time}
\label{sec:analysis}
We demonstrate the utility of \systemnameAPIBenchmark{} by showing interesting insights that we can learn from it regarding how APIs change over time. The analysis here is not meant to be exhaustive; indeed we leave many open directions of investigation and encourage the community to dive deeper using \systemnameAPIBenchmark{}.
 Our preliminary analysis goal is four-fold: (i) assess whether an ML API's predictions change over time, (ii) quantify how much accuracy improvements or declines are incurred due to ML API shifts, (iii) estimate to which direction prediction confidences of the ML APIs move, and (iv) understand how an ML API's gender and race biases evolve.  
 
\subsection{Findings on Classification APIs}
We first study the shifts of ML APIs designed for simple classification tasks, including facial emotion recognition, sentiment analysis, and spoken command detection. To quantify shifts on classification APIs, We adopt the following metrics:
\begin{itemize}
    \item \textbf{Prediction Overlap}. Prediction overlap measures how often an ML API's prediction on the same input  remains the same at different evaluation periods. 
    Formally, it can be expressed as 
    \begin{equation*}
        \PO(t_1,t_2) \triangleq \frac{1}{|D|}\sum_{(x,y)\in D}^{} \mathbbm{1}\left\{f(x,t_1)=f(x,t_2) \right\}
    \end{equation*}
    Here, $t_1$ and $t_2$ are two evaluation time periods. 
    $\PO=1$ indicates an ML API's predictions do not change, and $\PO=0$ means its predictions between $t_1$ and $t_2$ are completely different.  
    \item \textbf{Confidence Movement.} API shifts include both prediction and confidence score changes. 
    For a fixed data point, an ML API's prediction can remain the same, but its confidence may still move up and down over time. 
    To measure this, we use confidence movement  
        \begin{equation*}
        \ConfMove(t_1,t_2) \triangleq \frac{\sum_{(x,y)\in D}^{} \mathbbm{1}\left\{f(x,t_1)=f(x,t_2) \right\} \cdot \left[q(x,t_1) - q(x,t_2)\right] }  {\sum_{(x,y)\in D}^{} \mathbbm{1}\left\{f(x,t_1)=f(x,t_2) \right\}}
    \end{equation*}
    If $CM(t_1,t_2)>0$, then among all data points without prediction shifts, the evaluated ML API is more confident at time $t_1$ than at time $t_2$. 
    If $CM(t_1,t_2)<0$, then on average, the API's confidence is less confident at time $t_1$ than at time $t_2$.
    It is worth noting that many applications are sensitive to confidence  changes.
    For example, a customer review application  may trust an ML API's prediction if its confidence is larger than a threshold, and involve a human expert otherwise. 
    Even if all predictions stay the same, the confidence change over time may still mitigate or worsen the human expert's workload.  
    \item \textbf{Model Accuracy.}
    One of the most widely adopted ML API assessments is accuracy, i.e., how often the ML API makes the right prediction. Given a dataset $D$ and the label prediction   $f(\cdot,t)$ by an ML API  evaluated at time $t$, accuracy is simply
    \begin{equation*}
        \Acc(t) \triangleq \frac{1 }{|D|}  {\sum_{(x,y)\in D}^{} \mathbbm{1}\left\{f(x,t)=y \right\}}
    \end{equation*}  
    Thus, it is natural to quantify how the accuracy of an ML API changes over time. 
    
    \item 
    \textbf{Group Disparity.} Various metrics~\cite{barocas2017fairness,corbett2018measure,liu2018delayedfairness} have been proposed to quantify ML fairness. 
    In this paper, we adopt one common metric called  \textit{group disparity}~\cite{corbett2018measure}. Suppose the dataset $D$ is partitioned into $K$ groups $D_1,D_2,\cdots, D_K$ by some sensitive feature (e.g., gender or race). Then group disparity is 
    
    \begin{equation*}
        \GD(t) \triangleq \max_{i} \frac{1 }{|D_i|}  \left({\sum_{(x,y)\in D_i}^{} \mathbbm{1}\left\{f(x,t)=y \right\}}\right) - \min_{i} \frac{1 }{|D_i|}  \left({\sum_{(x,y)\in D_i}^{} \mathbbm{1}\left\{f(x,t)=y \right\}}\right)
    \end{equation*}  
In a nutshell, group disparity measures the accuracy difference between the most privileged group and the most disadvantaged group. 
Larger group disparity implies more unfairness, and $\GD(t)=0$ implies the API achieves perfect fairness at time $t$. 
 \end{itemize}

\paragraph{A case study on DIGIT.} We start with a case study on a spoken command recognition dataset, DIGIT~\cite{Dataset_Speech_DIGIT}. 
DIGIT contains 2,000 short utterances corresponding to digits from 0 to 9, and the task is to predict which number each utterance indicates.
We have evaluated three speech recognition APIs from IBM, Google, and Microsoft in year 2020, 2021, and 2022, separately. 
The utterances were spoken by people with US accent,  French accent, and German accent. 
Thus, we use accent as the sensitive feature to group the data instances and then measure the group disparity.

\begin{figure*}[t]
	\centering
	\vspace{-0mm}
	\includegraphics[width=0.95\linewidth]{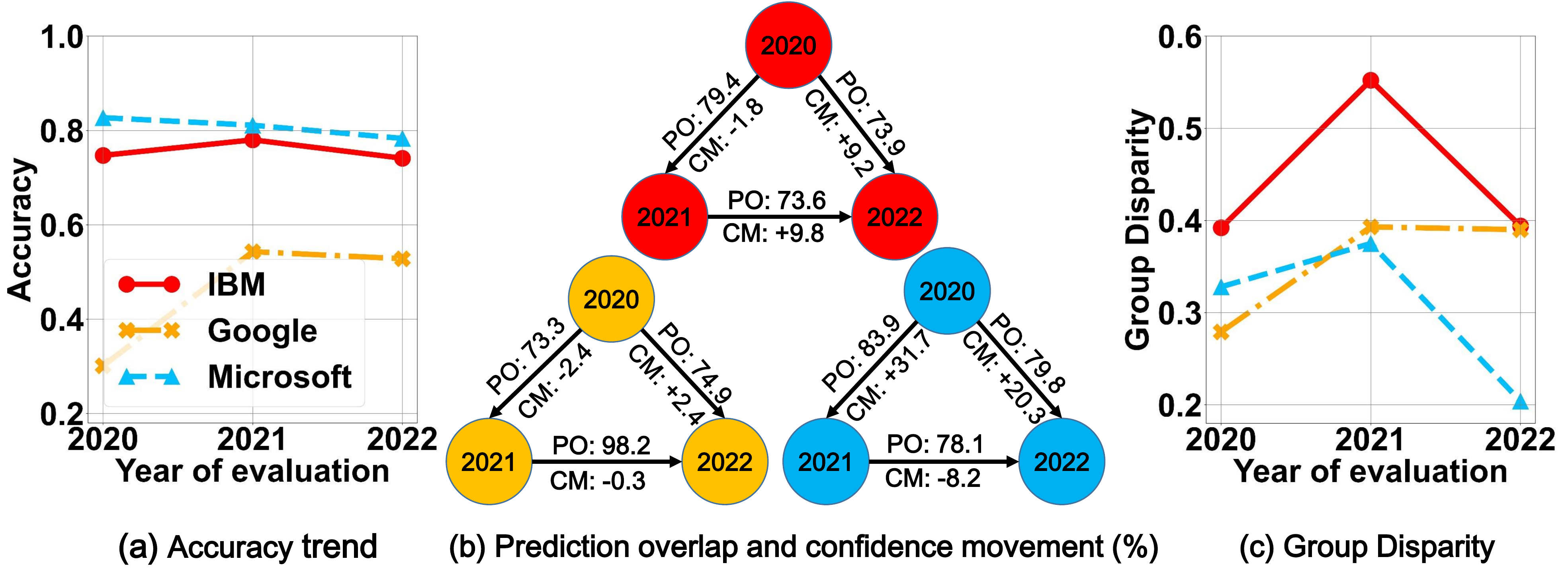}
	\vspace{-0mm}
	\caption{A case study on the dataset DIGIT.
(a): accuracy over time.	(b): prediction overlap and confidence movement of IBM, Google, and Microsoft APIs.  (c): group disparity with respect to speaker accent. 
	Overall, accuracy changes due to API shifts are notable, but the prediction changes are even more significant. For example, from 2020 to 2021, the accuracy of IBM API has increased by 4\% (see (a)), but  20.6\% predictions are actually  different (see (b)). 
	In addition, the confidence can move up by up to 31.7\% (Microsoft from 2020 to 2021 in (a))  while the prediction accuracy slightly drops (Microsoft in (b)). This calls for cautions in confidence-sensitive applications. It is also worth noting that large group disparity exists for  all evaluated APIs. 
	Interestingly, API update over time may either improve or hurt overall accuracy as well as group fairness.}
	\label{fig:APIBenchmark:casestudy1}
\end{figure*}

As shown in Figure \ref{fig:APIBenchmark:casestudy1}, there are many interesting observations in this case study.
First, the accuracy changes are substantial: for example, as shown in Figure \ref{fig:APIBenchmark:casestudy1}(a), Google API's accuracy increased by 20\% from 2020 to 2021.
The prediction changes are even more significant: 
For example, from 2020 to 2021, IBM API's accuracy rose by 4\% (see Figure \ref{fig:APIBenchmark:casestudy1}(a), but  1-79.4\%=20.6\% predictions by IBM API were changed (see Figure \ref{fig:APIBenchmark:casestudy1}(b)). This is perhaps because while some mistakes were fixed by the API update, some utterances previously correctly predicted may be predicted incorrectly by the updated version. 
Even when the predictions remain steady, the confidence score can still significantly move up or down. 
For example, the confidence produced by Microsoft API moved up by 31.7\% from 2020 to 2021 (as shown in Figure \ref{fig:APIBenchmark:casestudy1}(b)). 
Yet, its accuracy dropped by 1.5\% (as shown in Figure \ref{fig:APIBenchmark:casestudy1}(a)).
This raises cautions in downstream applications that rely on confidence scores.
It is also worth noting that group disparity exists for all evaluated APIs. 
In 2020, perhaps surprisingly, Google API's accuracy is the lowest but its disparity is also the smallest.
Model update may either improve or hurt accuracy and group disparity. For example, Google API's accuracy is improved over time but its bias towards non-native accents is also worsen. 

\begin{figure*}[t]
	\centering
	\vspace{-0mm}
	\includegraphics[width=0.999\linewidth]{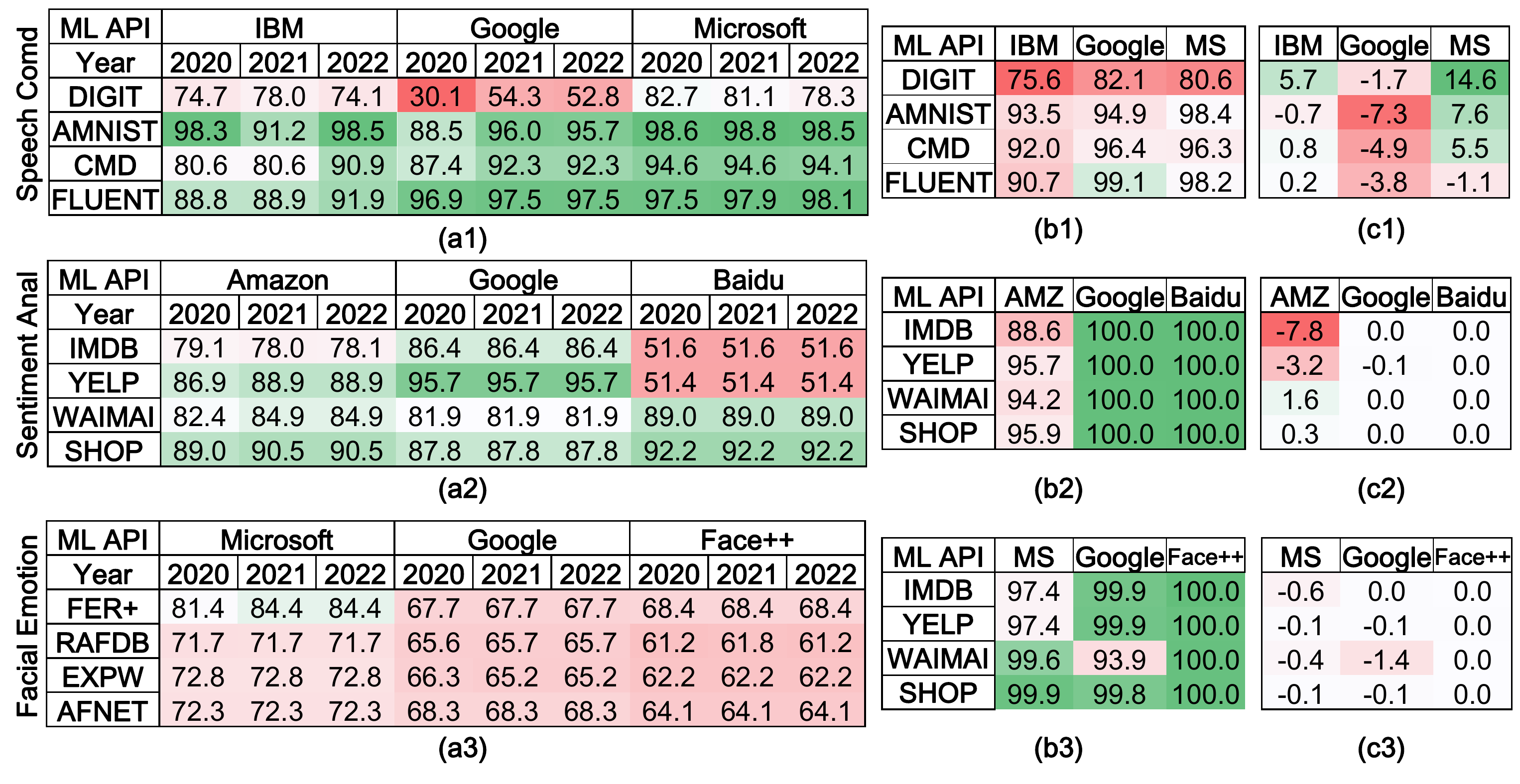}
	\vspace{-0mm}
	\caption{Summary on classification API shifts from 2020 to 2022. Tables on row 1, 2 and 3  correspond to spoken command recognition, sentiment analysis, and facial emotion recognition, respectively. (a1)-(a3): Accuracy of each year.
	(b1)-(b3): Average prediction overlap. (c1)-(c3): Average confidence movement. Units: \%.
	Red and green  indicate low and high values, respectively. 
	The accuracy changes  exhibit various patterns overall, while the API shifts are also diverse: all spoken command recognition APIs' predictions have been changed significantly during the past years, while significant changes exist for only one third of the APIs for the other two tasks. Confidence movements are also interesting. 
	For example, Google  API for spoken command recognition tends to be  less confident (c1), while Amazon sentiment API is more confident on Chinese texts but less on English texts (c2).  }
	\label{fig:APIBenchmark:full_class_API}
\end{figure*}

\paragraph{Diverse API shifts across multiple classification tasks.} Next we study API shifts across different tasks and datasets. 
For each API dataset pair, we calculate the  prediction overlap and confidence movement between each evaluation time pair (2020--2021, 2020--2022, 2021--2022) and then report the results averaged over all time pairs.  We also measure and compare its accuracy for each year. The results are shown in Figure \ref{fig:APIBenchmark:full_class_API}. 

Several interesting findings exist. First, small accuracy changes may be the result of large prediction shifts, i.e., small  prediction overlaps. 
For example, about 10\% predictions made by Amazon sentiment analysis API on IMDB (as shown in Figure \ref{fig:APIBenchmark:full_class_API}(b1)) have changed, but its accuracy only changes by about 1\% (Figure \ref{fig:APIBenchmark:full_class_API}(a1)). 
Similarly, a 3\% prediction  difference exists for Microsoft API on RAFDB (Figure \ref{fig:APIBenchmark:full_class_API}(b3)) while there is almost no change in its accuracy (Figure \ref{fig:APIBenchmark:full_class_API}(a3)).
This indicates general phenomena in API shifts: many API updates fix certain errors but also make additional mistakes.   
Next, we note that the API shifts are diverse. For spoken command recognition, all evaluated APIs' predictions are changed significantly (Figure \ref{fig:APIBenchmark:full_class_API}(b1)). 
However, the shifts in APIs for facial emotion recognition is almost negligible (Figure \ref{fig:APIBenchmark:full_class_API}(b3)). 
This implies that different APIs may be updated in a different rate and thus detecting whether a shift may have happened is useful.
Moreover, different APIs' confidence movements are not similar. 
Sometimes an ML API tends to be more and more conservative: for example, the average confidences of Google API for spoken command recognition have dropped notably for all evaluated datasets. 
Sometimes an ML API becomes more and more confident: 
for example, Microsoft API for spoken recognition has increased its confidence over time on three out of four datasets ( Figure \ref{fig:APIBenchmark:full_class_API}(c1)).  
More interestingly, its confidence may also depend on a dataset's property: as shown in Figure \ref{fig:APIBenchmark:full_class_API}(c2), Amazon sentiment  analysis API tends to be less confident on Chinese texts (WAIMAI and SHOP) but more confident on English texts (IMDB and YELP). 
Understanding how the confidence moves may help decision making in confidence-sensitive applications. 
We provide additional group disparity analysis in the appendix.

\subsection{Findings on  Structured Prediction APIs}
Next we turn to the structured prediction APIs. 
Similar to standard classification APIs, we use prediction overlap to measure prediction changes due to shifts of structured predictions APIs. For each data instance, we use the average of all predicted labels' confidences as an overall confidence, and then still apply confidence movement to quantify how an API's confidence shifts over time. 
To measure structured prediction API's performance, we adopt the standard multi-label accuracy
\begin{equation*}
        \MultiAcc(t) \triangleq \frac{1 }{|D|} \sum_{(x,y)\in D}^{} \frac{|f(x,t) \cap y|}{|f(x,t) \cup y|} 
        \end{equation*}
Finally, we keep using group disparity to evaluate fairness of an ML API, but replace the 0-1 loss $\mathbbm{1}\{f(x,t)=y\}$ by the Jaccard similarity $\frac{|f(x,t) \cap y|}{|f(x,t) \cup y|}$.

\begin{figure*}[t]
	\centering
	\vspace{-0mm}
	\includegraphics[width=0.93\linewidth]{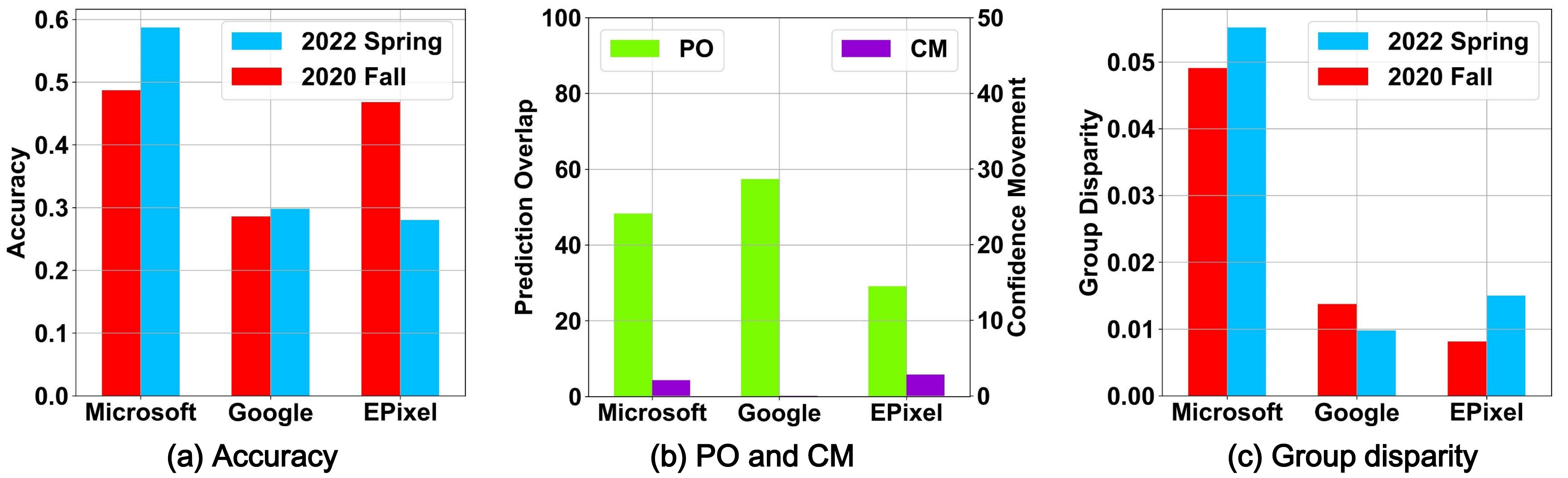}
	\vspace{-0mm}
	\caption{A case study on the  dataset COCO.
	(a): accuracy over time.
	(b): prediction overlap (\%)  and confidence movement (\%). (c): group disparity with respect to gender. 
	Here, the accuracy change is quite significant.
	E.g.,  EPixel API update leads to  20\% accuracy drop (as shown in  (a)).  Prediction shifts are also large:  prediction overlap can be less than 30\% (as shown in (b)).  
	The confidence movement is relatively small. 
 It is also worth noting that high accuracy does not imply better fairness. In fact,   Microsoft API's accuracy is the highest, but its group disparity is also the largest (c).}
	\label{fig:APIBenchmark:casestudy_coco}
\end{figure*}

\paragraph{A case study on COCO.}
We start with a case study on the dataset COCO. 
COCO contains more than a hundred thousand images, and the goal is to determine if one or more objects from 80 categories show up in each image. 
We have evaluated three APIs from Microsoft, Google, and EPixel, respectively.  
To measure group disparity, we adopt the gender labels~\cite{zhao2021cocometa} for a subset of COCO which contains a person, and then calculate the group disparity on this subset for all evaluated ML APIs.
The results are summarized in Figure \ref{fig:APIBenchmark:casestudy_coco}.

Our first observation is that the accuracy shift can be quite large, leading to an ``accuracy cross''. 
As shown in Figure \ref{fig:APIBenchmark:casestudy_coco}(a), EPixel API's accuracy drops by more than 20\% while Google API's accuracy increases by 1\%. 
Consequentially, Google API becomes more accurate than EPixel, while the latter was more accurate in 2020. This implies that API shifts can be impactful in business decision making such as picking which ML API to use.
In addition, the prediction shifts are much larger than those for simple classification APIs. 
For example, prediction overlap for EPixel is less than 30\%, meaning that 70\% of the predictions are different than before (as shown in Figure \ref{fig:APIBenchmark:casestudy_coco}(b)). On the other hand, the confidence movement is relatively small: as shown in Figure \ref{fig:APIBenchmark:casestudy_coco}(b), no API's confidence movement is larger than 3\%.
It is also worth noting that high accuracy does not imply better fairness necessarily. In fact,   Microsoft API's accuracy is the highest, but its group disparity is also the largest.
As shown in Figure \ref{fig:APIBenchmark:casestudy_coco}(a) and (c), API shifts may improve the accuracy but simultaneously amplify the group disparity: Microsoft API's accuracy increases by 10\% but its group disparity is also enlarged.

\begin{figure*}[t]
	\centering
	\vspace{-0mm}
	\includegraphics[width=0.95\linewidth]{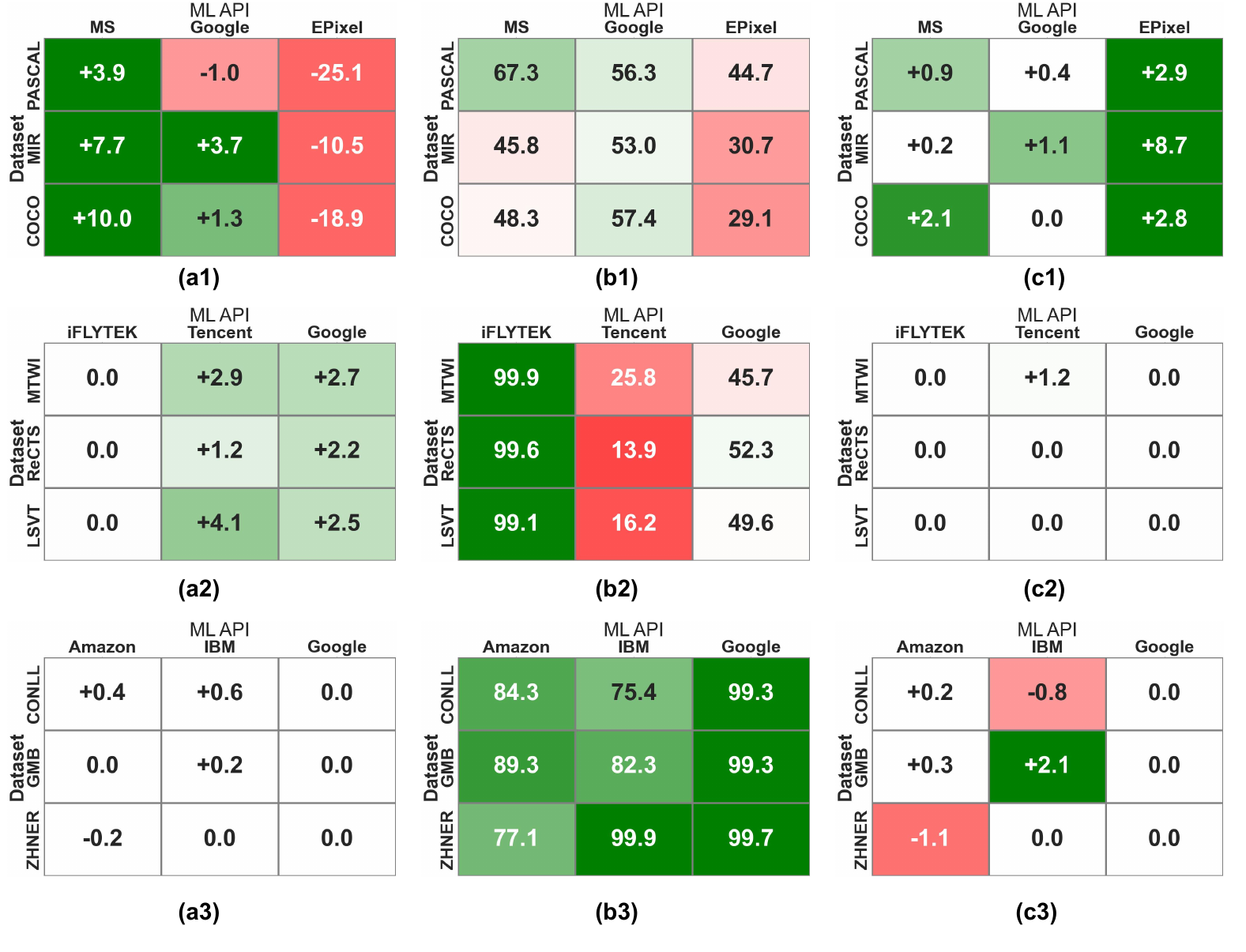}
	\vspace{-0mm}
	\caption{Summary on structured prediction API shifts.
	Row 1, 2, and 3 correspond to multi-label image classification, scene text recognition and named entity recognition.
	The left, middle, and right column correspond to
 accuracy changes (\%),  prediction overlaps (\%), and confidence movements (\%), respectively, between 2020 and 2022. The accuracy change is large for multi-label image classification but relatively small for the other two tasks. 
Predicted labels change notably for many of the evaluated ML APIs.  
The confidence movement is relatively small, though. }
	\label{fig:APIBenchmark:MLCfullanalysis}
\end{figure*}

\paragraph{Various API shift patterns across structured prediction tasks.} 
Finally, we dive deeply into various API shift patterns for more structured prediction tasks. 
The prediction overlaps, confidence movements, and accuracy changes for 27 API-dataset pairs are summarized in Figure \ref{fig:APIBenchmark:MLCfullanalysis}.

There are several interesting observations.
First, the accuracy changes are significant for multi-label image classification but relatively small for the other two tasks, as shown in Figure \ref{fig:APIBenchmark:MLCfullanalysis}(a1)-(a3). 
However, API shifts for structured predictions  are more common than classification tasks.
In fact, as shown in Figure \ref{fig:APIBenchmark:MLCfullanalysis}(b1)-(b3), prediction changes occur for almost all ML APIs. 
The magnitudes of the shifts are also larger. 
This is perhaps because structured prediction is more sensitive to model updates than those for  classifications. 
The confidence movement is relatively small though. 
Note that confidence movements do not always reflect the APIs' performance changes. 
For instance, EPixel API's confidence increases on all evaluated datasets, but its accuracy actually drops. This is probably because EPixel's update removes a label due to low confidence but this label was  part of the true label set.  
Detecting, estimating, and explaining such phenomena is needed for robustly adopting ML APIs.

\section{Additional Discussions and Maintenance Plans}
\begin{figure*}[t]
	\centering
	\vspace{-0mm}
	\includegraphics[width=0.97\linewidth]{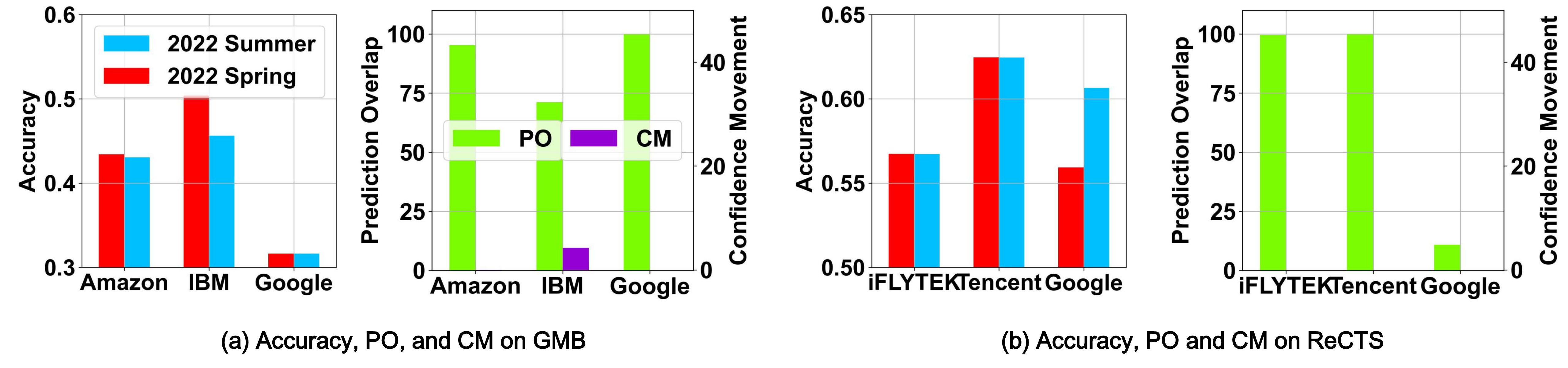}
	\vspace{-0mm}
	\caption{API Shifts within 6 months. (a) and (b) correspond to the GMB and ReCTS datasets, respectively. Overall, significant prediction and accuracy occurred in 3 out 6 ML APIs.}
	\label{fig:APIBenchmark:6monthshift}
\end{figure*}

\paragraph{More frequent evaluations.} ML APIs are increasingly growing and  updated frequently. Thus, we plan to enrich our database by continuously evaluating ML APIs more frequently, i.e., every 6 months. 
As of 2022 August, we have collected additional  predictions of all structured prediction APIs.
As shown in Figure \ref{fig:APIBenchmark:6monthshift}, significant prediction changes already occurred in 6 months.
For example, the accuracy of IBM named entity API on the GMB dataset dropped from 50\% (March 2022) to 45\% (August 2022).
Those newly collections have been added to our database. More details can be found in the appendix.

\paragraph{Comparison with open-source ML models.}
As a baseline, we have also measured the performance of several open source ML models on all  datasets.
As shown in Table \ref{tab:APIShiftBenchmark:Baselines}, the open source ML models' performance varies across different datasets, and can be sometimes better than that of commercial APIs. 
This further emphasizes the importance of monitoring commercial APIs' performance. 

\begin{table}[htbp]
  \centering
  \small
  \caption{Performance of open source models on the evaluated datasets. For some tasks, open source models' performance can be even better than  that of the commercial APIs.}
    \begin{tabular}{|c||c|c|c|c||c|c|c|}
    \hline
    Task  & \multicolumn{4}{c||}{Speech Recognition} & \multicolumn{3}{c|}{Multi-label Image Classification} \bigstrut\\
    \hline
    Open source model & \multicolumn{4}{c||}{DeepSpeech~\cite{DeepSpeech_ICML16}} & \multicolumn{3}{c|}{SSD~\cite{SSD_MIC_github}} \bigstrut\\
    \hline
    Dataset & DIGIT & AMNIST & CMD   & FLUENT & PASCAL & MIR   & COCO \bigstrut\\
    \hline
    Performance & 0.60  & 0.92  & 0.80  & 0.87  & 0.64  & 0.25 & 0.40  \bigstrut\\
    \hline
    \hline

    Task  & \multicolumn{4}{c||}{Sentiment Analysis} & \multicolumn{3}{c|}{Scene Text Recognition} \bigstrut\\
    \hline
    Open source model & \multicolumn{4}{c||}{Vader~\cite{VanderICWSM2014}}    & \multicolumn{3}{c|}{PP-OCR~\cite{PP_OCRPaper2020pp}} \bigstrut\\
    \hline
    Dataset & IMDB  & YELP  & WAIMAI & SHOP  & MTWI  & ReCTS & LSVT \bigstrut\\
    \hline
    Performance & 0.69  & 0.75  & 0.64  & 0.78  & 0.63  & 0.51  & 0.47  \bigstrut\\
    \hline
    \hline
    
    Task  & \multicolumn{4}{c||}{Facial Emotion Recognition} & \multicolumn{3}{c|}{Named Entity Recognition} \bigstrut\\
    \hline
    Open source model & \multicolumn{4}{c||}{A convolution neural network~\cite{FER_github}}      & \multicolumn{3}{c|}{Spacy~\cite{Spacy_github}} \bigstrut\\
    \hline
    Dataset & FER+  & RAFDB & EXPW  & AFNET & CONLL & GMB   & ZHNER \bigstrut\\
    \hline
    Performance & 0.77  & 0.60  & 0.56  & 0.64  & 0.53  & 0.55  & 0.63  \bigstrut\\
    \hline
    \end{tabular}%
  \label{tab:APIShiftBenchmark:Baselines}%
\end{table}%

\paragraph{Expansion of Datasets and ML APIs.}
Part of the future plan is to expand the scope of datasets and ML APIs. 
To allow this, we plan to solicit needs from the ML communities: a poll panel will be created on our website, and ML researchers, engineers, and domain experts are all welcome  to vote for which ML APIs and which datasets to include in our database. We will periodically update the database based on the community's feedback.

\section{Conclusions and Open Questions}
\label{sec:conclusions}
ML APIs play an increasingly important role in real world ML adoptions and applications, but there are only a limited number of papers studying the properties and dynamics of these commercial APIs. 
In this paper, we introduce \systemnameAPIBenchmark{}, a large scale dataset consisting of samples from various tasks annotated by ML APIs over multiple years.
Our analysis on \systemnameAPIBenchmark{} shows interesting findings, including large price gaps among APIs for the same task,  prevalent ML API shifts between 2020 and 2022, diverse performance differences between API venders, and consistent subgroup performance disparity. And this is just scratching the surface. 
\systemnameAPIBenchmark{} enables many interesting questions to be studied in the ML marketplaces. A few examples include:
\begin{itemize}
    \item How to determine which API or combination of APIs to use for any given application? \systemnameAPIBenchmark{} can serve as a testbed to evaluate and compare different API calling strategies.
    \item How to  perform unsupervised or semi-supervised performance estimation under ML API shifts? This is useful for practical ML monitoring but not possible without a detailed ML API benchmark over time provided by  \systemnameAPIBenchmark{}. 
    
    \item  Generally, how to estimate performance shifts when both ML APIs and data distributions shift? 
    
    \item How to explain the performance gap due to ML API shifts? More fine-grained understanding of how the API's prediction behavior changes over time would be useful for practitioners. 
    
\end{itemize}
These and other open questions enabled by \systemnameAPIBenchmark{} are increasingly critical with the growth of ML-as-a-service.
 \systemnameAPIBenchmark{} can greatly stimulate more research on ML marketplace.
 All of the data in \systemnameAPIBenchmark{} is openly available at \url{https://github.com/lchen001/HAPI}.

\newpage
{
\small

\bibliography{MLService}
\bibliographystyle{plain}
}

\section*{Checklist}

\begin{enumerate}

\item For all authors...
\begin{enumerate}
  \item Do the main claims made in the abstract and introduction accurately reflect the paper's contributions and scope?
    \answerYes{See Abstract and Section~\ref{sec:introduction}.}
  \item Did you describe the limitations of your work?
    \answerYes{See Section~\ref{sec:conclusions}.}
  \item Did you discuss any potential negative societal impacts of your work?
    \answerYes{See Section~\ref{sec:conclusions}.}
  \item Have you read the ethics review guidelines and ensured that your paper conforms to them?
    \answerYes{}
\end{enumerate}

\item If you are including theoretical results...
\begin{enumerate}
  \item Did you state the full set of assumptions of all theoretical results?
    \answerNA{}
	\item Did you include complete proofs of all theoretical results?
    \answerNA{}
\end{enumerate}

\item If you ran experiments (e.g. for benchmarks)...
\begin{enumerate}
  \item Did you include the code, data, and instructions needed to reproduce the main experimental results (either in the supplemental material or as a URL)?
    \answerYes{See \url{https://github.com/lchen001/HAPI/} and the supplement material.}
  \item Did you specify all the training details (e.g., data splits, hyperparameters, how they were chosen)?
    \answerNA{}
	\item Did you report error bars (e.g., with respect to the random seed after running experiments multiple times)?
    \answerNA{}
	\item Did you include the total amount of compute and the type of resources used (e.g., type of GPUs, internal cluster, or cloud provider)?
    \answerNA{}
\end{enumerate}

\item If you are using existing assets (e.g., code, data, models) or curating/releasing new assets...
\begin{enumerate}
  \item If your work uses existing assets, did you cite the creators?
    \answerYes{See Section~\ref{sec:construction}.}
  \item Did you mention the license of the assets?
    \answerNA{}
  \item Did you include any new assets either in the supplemental material or as a URL?
    \answerYes{See \url{https://github.com/lchen001/HAPI/}.}
  \item Did you discuss whether and how consent was obtained from people whose data you're using/curating?
    \answerNA{}
  \item Did you discuss whether the data you are using/curating contains personally identifiable information or offensive content?
    \answerNA{}
\end{enumerate}

\item If you used crowdsourcing or conducted research with human subjects...
\begin{enumerate}
  \item Did you include the full text of instructions given to participants and screenshots, if applicable?
    \answerNA{}
  \item Did you describe any potential participant risks, with links to Institutional Review Board (IRB) approvals, if applicable?
    \answerNA{}
  \item Did you include the estimated hourly wage paid to participants and the total amount spent on participant compensation?
    \answerNA{}
\end{enumerate}

\end{enumerate}


\appendix

\section*{Acknowledgement}
This project is supported in part by NSF CCF 1763191, NSF CAREER AWARD
1651570 and NSF CAREER AWARD 1942926.
There is no industrial funding.
We appreciate the anonymous reviewers for their invaluable feedback. 
\newpage
\section*{Supplementary materials}
The supplementary materials include additional details of \systemnameAPIBenchmark{} and extra model shift study.

\section{Additional Details of \systemnameAPIBenchmark}
In this section, we provide additional details of the constructed dataset \systemnameAPIBenchmark{}, including  motivation,  composition, collection process, preprocessing and cleaning, uses,  distribution, and maintenance.

\subsection{Motivation}
\systemnameAPIBenchmark{} was created to enable research on ML APIs.
This includes but is not limited to, for example, determing which API or combination of APIs to use for different user data or applications as well as budget constraints, estimating how much performance has changed due to API shifts, and  explaining  the performance gap due to ML API  shifts.

\subsection{Composition, and Collection Process}
Each instance in \systemnameAPIBenchmark{} consists of a query input for an API (e.g., an image or text) along with the API's output prediction/annotation and confidence scores.
For example, one instance could be an image from the image dataset COCO~\cite{Dataset_COCO_2014}, and \{\textit{(person, 0.9), (sports ball, 0.78),
(tennis racket, 0.45)}\}, the associated annotation by Microsoft API. This means Microsoft API predicts three labels, \textit{person, sports}, and \textit{tennis racket}, with confidence scores \textit{0.9, 0.78}, and \textit{0.45}, respectively.   

The query inputs were collected from 21 datasets for 6 different tasks. 
For SCR, four datasets were used: 
DIGIT~\cite{Dataset_Speech_DIGIT}, AMNIST~\cite{Dataset_Speech_AudioMNIST_becker2018interpreting}, CMD~\cite{Dataset_Speech_GoogleCommand}, 
and FLUENT~\cite{Dataset_Speech_Fluent_LugoschRITB19}. 
The sampling rate is 8 kHz, 48 kHz, 16 kHz, and 16 kHz, respectively.  Each utterance is a spoken digit (i.e., 0-9) in DIGIT and AMNIST and a short command from a total of 30 commands (such
as “go”, “left”, “right”, “up”, and “down”) or white noises in CMD. 
In FLUENT, the
commands are typically a phrase (e.g., “turn on the light” or “turn down the music”) from a total of 248 phrases. 
Four text datasets were used for SA: YELP~\cite{Dataset_SEntiment_YELP}, IMDB~\cite{Dataset_SEntiment_IMDB_ACL_HLT2011}, SHOP~\cite{Dataset_SENTIMENT_SHOP}, and WAIMAI~\cite{Dataset_SENTIMENT_WAIMAI}. YELP and IMDB are
both in English while WAIMAI and SHOP are in Chinese. 
FER+~\cite{dataset_FERP_BarsoumZCZ16}, RAFDB~\cite{Dataset_FAFDB_li2017reliable},
EXPW~\cite{Dataset_EXPW_SOCIALRELATION_ICCV2015}, and AFNET~\cite{Dataset_AFFECTNET_MollahosseiniHM19} were used for FER task.
The emotion labels were anger,  disgusting, fear,  happy, sad, surprise,  and natural. 

Three datasets were used for MIC: PASCAL~\cite{Dataset_Pascal_2015}, MIR~\cite{Dataset_MIR_2008}, and COCO~\cite{Dataset_COCO_2014}.
There are 20 and 80 distinct labels in PASCAL and COCO, respectively. 
MIR contains 25 unique labels, and we removed the label ``night'' as it is not in the label set of any ML APIs.
For STR, we adopted   MTWI~\cite{Dataset_MTWI_2018}, ReCTS~\cite{Dataset_ReCTS_2019}, and LSVT~\cite{Dataset_LSVT_2019}, three datasets containing real world images with Chinese texts.
Finally, for NER task, we used CONLL~\cite{Dataset_CONLL_2003}, ZHNER~\cite{ZHNER_dataset_github}, and GMB~\cite{Dataset_GMB_2013}.
CONLL and GMB contain both English texts while ZHNER is a Chinese text dataset. 

For each instance  in those datasets, we have evaluated the prediction from the mainstream ML APIs from 2020 to 2022.
\systemnameAPIBenchmark{} was collected from 2020 to 2022.
For classification tasks, the predictions/annotations of each API were collected in the spring of 2020, 2021, and 2022.
For structured predictions, all APIs' predictions were collected in fall 2020 and spring 2022, separately. 
The details  can be found in Table \ref{tab:APIBenchmark:MLAPIs}.

\subsection{Preprocessing and Cleaning}
This includes both (i) preprocessing on the original inputs to the ML APIs and (ii) cleaning of the ML APIs' outputs.
The preprocessing on the original inputs is as follows. On FLUENT, all 248 unique phrases were mapped to 31 unique commands as provided in the original source~\cite{Dataset_Speech_Fluent_LugoschRITB19}. 
The original labels in YELP are user ratings (1,2,3,4, and 5).
1 and 2 were  transformed to negative; 3, 4, and 5 were mapped to positive. 
IMDB, WAIMAI and SHOP contain polarized review labels and thus we directly used those labels. 
As a result, classification on all SA datasets is  a binary task.
We used a sampled version of YELP:  10,000 text paragraphs with label positive and negative separately were randomly drawn from the original YELP dataset. 
The original IMDB dataset has been partitioned into training and testing splits, and thus we used its testing split, including 25,000 text paragraphs.
All instances in WAIMAI and SHOP were used. 
The facial images in FER+ was the same as the FER dataset from the ICML 2013 Workshop on Challenges in Representation. 
A training and testing  split and regenerated labels are provided in FER+. 
We adopted the testing split with the regenerated labels.
RAFDB and AFNET contain images for both basic emotions (anger, fear, disgusting, happy, sad, surprise, and natural) and compound emotions. 
We only evaluated ML APIs on images for basic emotions, as all evaluated ML APIs focus on basic emotions. 
Different from FER+, RAFDB, and AFNET, an image in EXPW may contain multiple faces.
Thus, the labels include both the bounding box and the labelling workers' confidence. 
Thus, we extracted aligned faces as ML APIs' inputs by enlarging by 10\% and then cropping the provided face bounding boxes whose  confidence scores are larger than 0.6.

Less preprocessing was performed for structured prediction datasets. For MIC, we directly sent all raw images to the ML APIs.
A diverse collection of images is included for STR: images for advertising sales forms MTWI, while most images in ReCTS are photos taken on sing boards. LSVT's iamges are typically street view images. While all images in MTWI and ReCTS are fully annotated, LSVT contains both fully and partially annotated images. 
\systemnameAPIBenchmark{} only considers the images with full annotations as inputs to ML APIs.
For NER datasets, all samples were included in \systemnameAPIBenchmark{}.
Yet, we only focused on three widely used types of entities: person, location, and organization. 
 
Different ML APIs may use different label sets for the same tasks. 
For example, both ``disgust'' and  ``disgusting'' may be returned by different ML APIs to refer to the same  facial emotion.
Thus, label alignment is needed. For classification tasks, we manually matched each API's predicted labels to a unique number. 
For example, for FER datasets, both ``happy'' and ``happiness'' were mapped to label 3, and label 4 corresponded to ``sad'', ``sadness'', and ``unhappiness''.
For MIC with less than 100 unique labels, we were able to create the label maps manually too.  
On STR datasets, predictions (i) that are  within 0-9 or (ii) whose unicode is in the range of  u4e00-u9fff are maintained. 
For NER, we also manually mapped each API's entity type to a universal type. For example, ``people'' and ``human'' are both mapped to ``person''.

\subsection{Uses, Distribution, and Maintenance}
\systemnameAPIBenchmark{} has been tested and used in this paper at the time of publication.
It can be used in any research related to ML prediction APIs or marketplaces, too. 
We will also maintain an incomplete list of which papers or projects have been developed on top of \systemnameAPIBenchmark{}.

The dataset is publicly available on the internet. 
The dataset is distributed on Lingjiao Chen's website: \url{https://github.com/lchen001/HAPI} under Apache License 2.0.
It was first released in 2022.
The dataset will be maintained by Lingjiao Chen and other authors of this paper. 
In addition, \systemnameAPIBenchmark{} will also be updated every few months to include up-to-date predictions from the mainstream ML APIs as well as emerging ML APIs. 
All the details and updates can be accessed on 
\url{https://github.com/lchen001/HAPI}.

\section{Extra Model Shift Study}

\begin{figure}[t] \centering
\begin{subfigure}[Accuracy on AMNIST ]{\label{fig:sample_a}\includegraphics[width=0.49\linewidth]{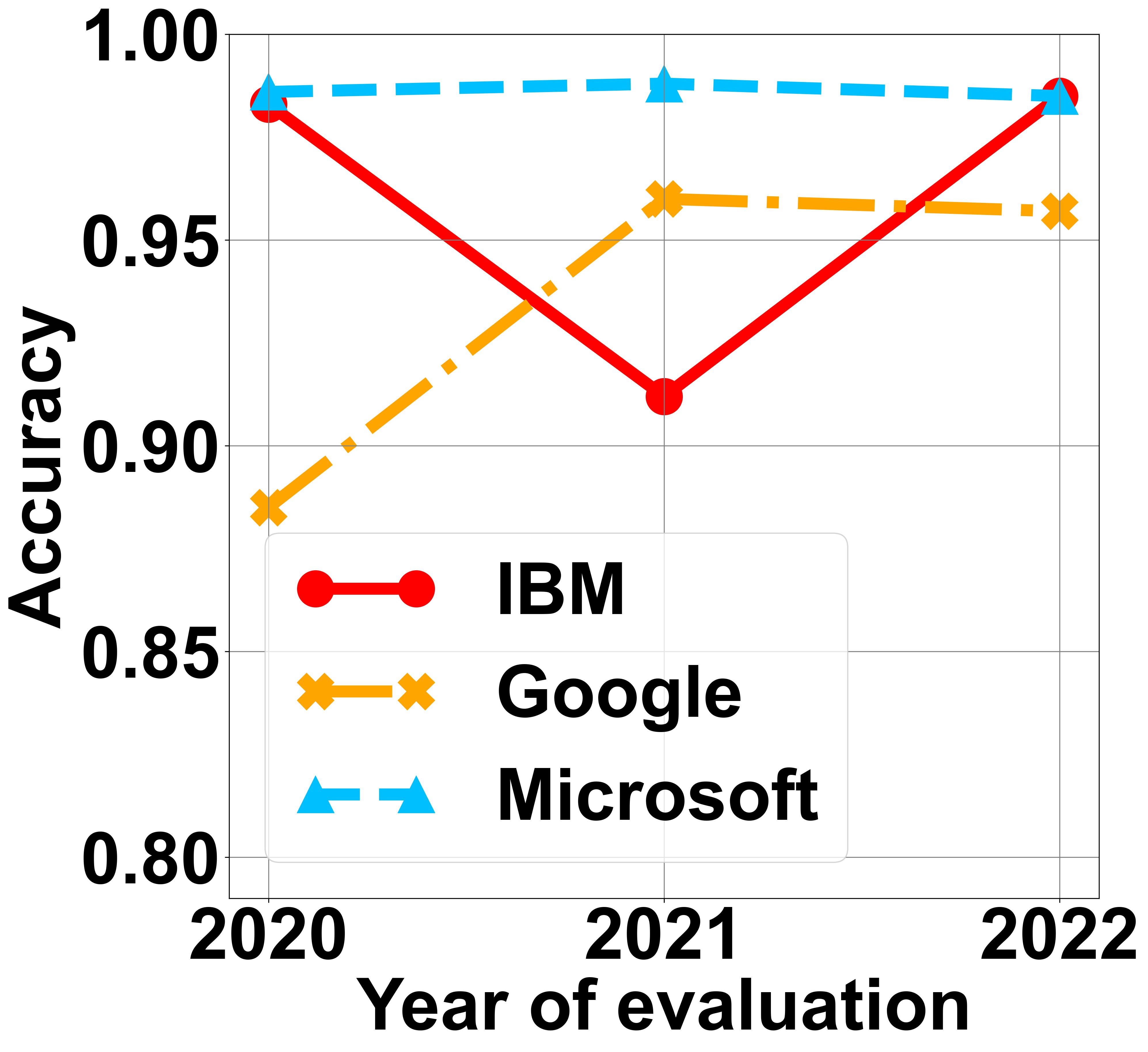}}
\end{subfigure}
\begin{subfigure}[Group Disparity on AMNIST]{\label{sample_b}\includegraphics[width=0.49\linewidth]{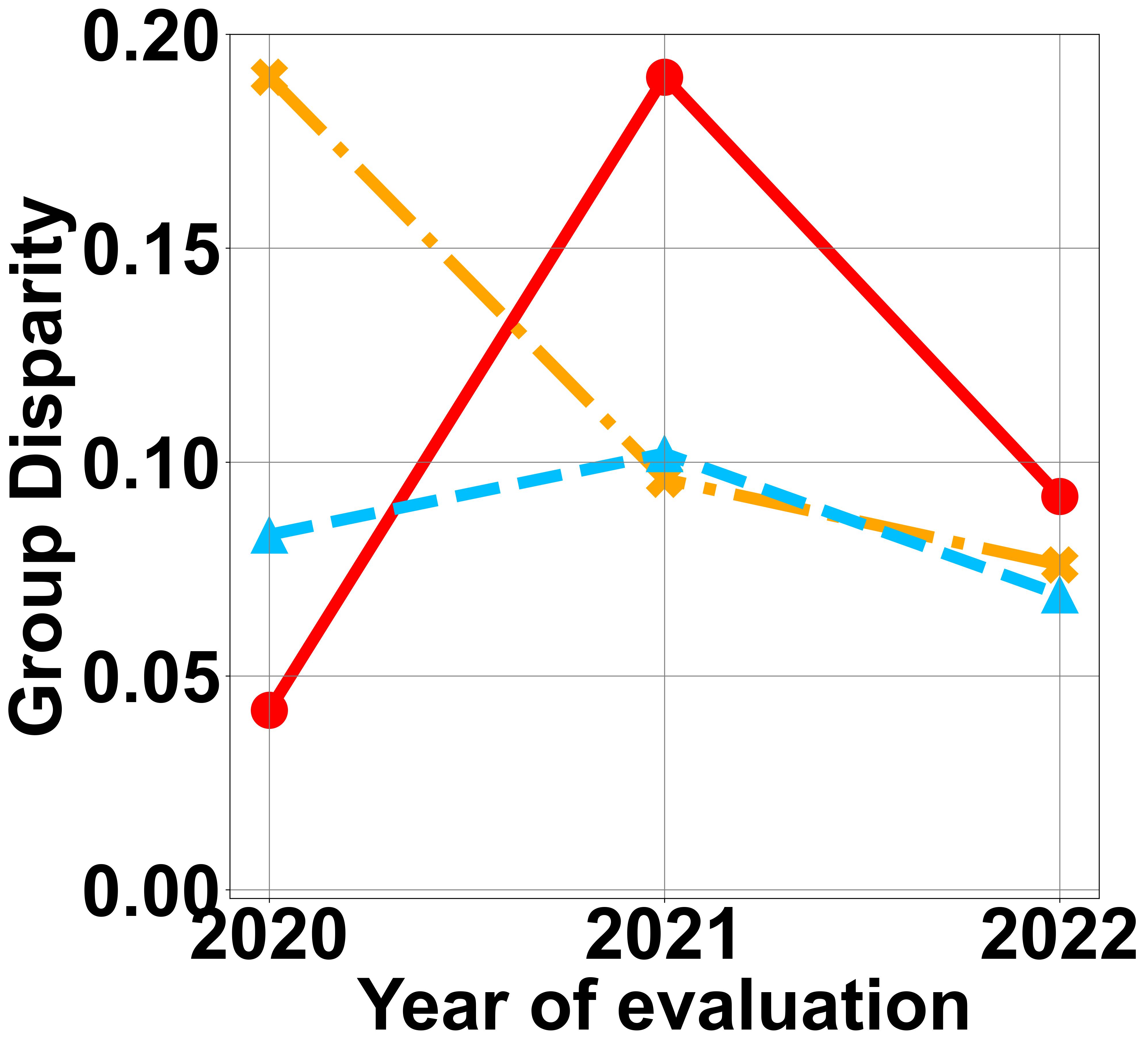}}
\end{subfigure}
\begin{subfigure}[Accuracy on RAFDB]{\label{sample_c}\includegraphics[width=0.49\linewidth]{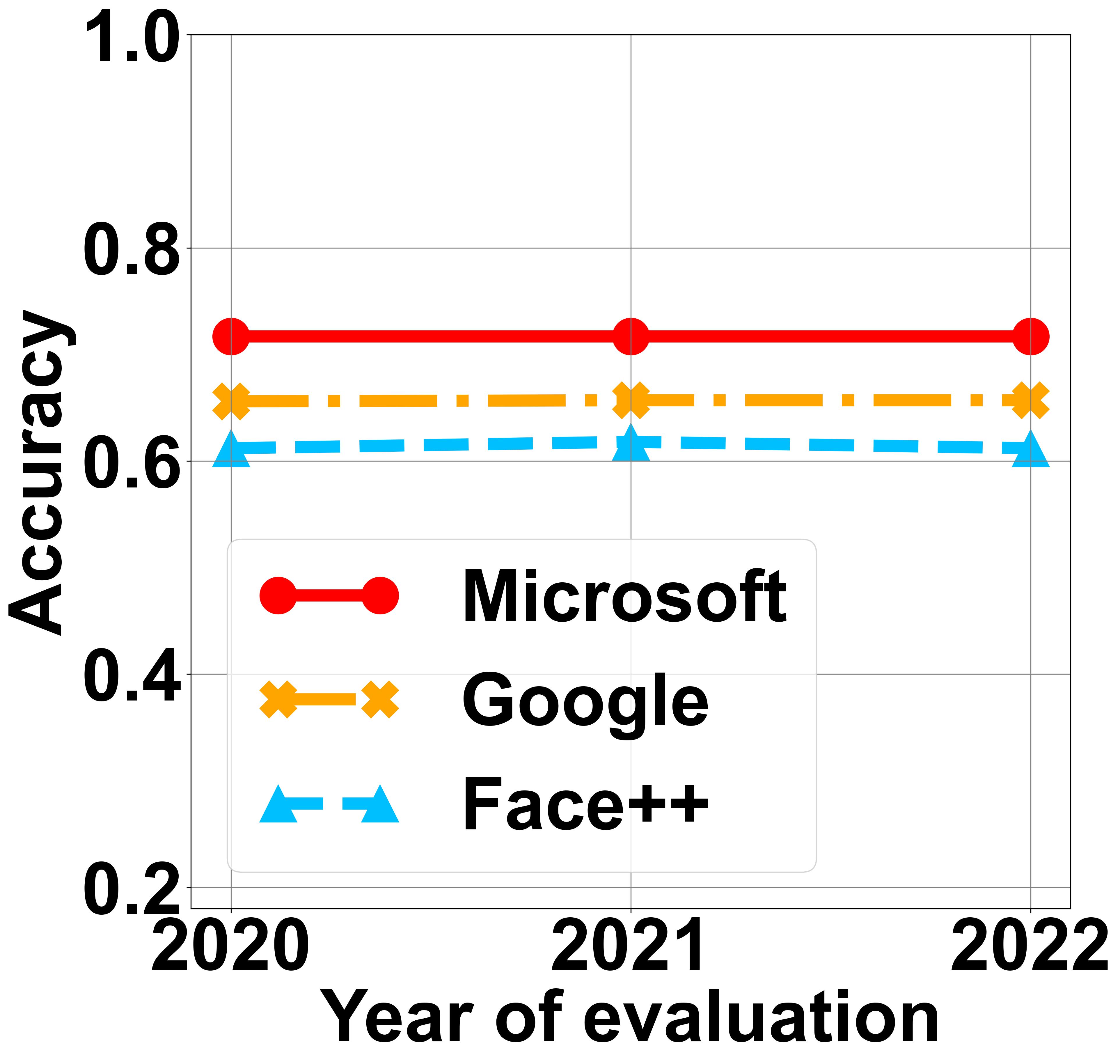}}
\end{subfigure}
\begin{subfigure}[Group Disparity on RAFDB]{\label{sample_c}\includegraphics[width=0.49\linewidth]{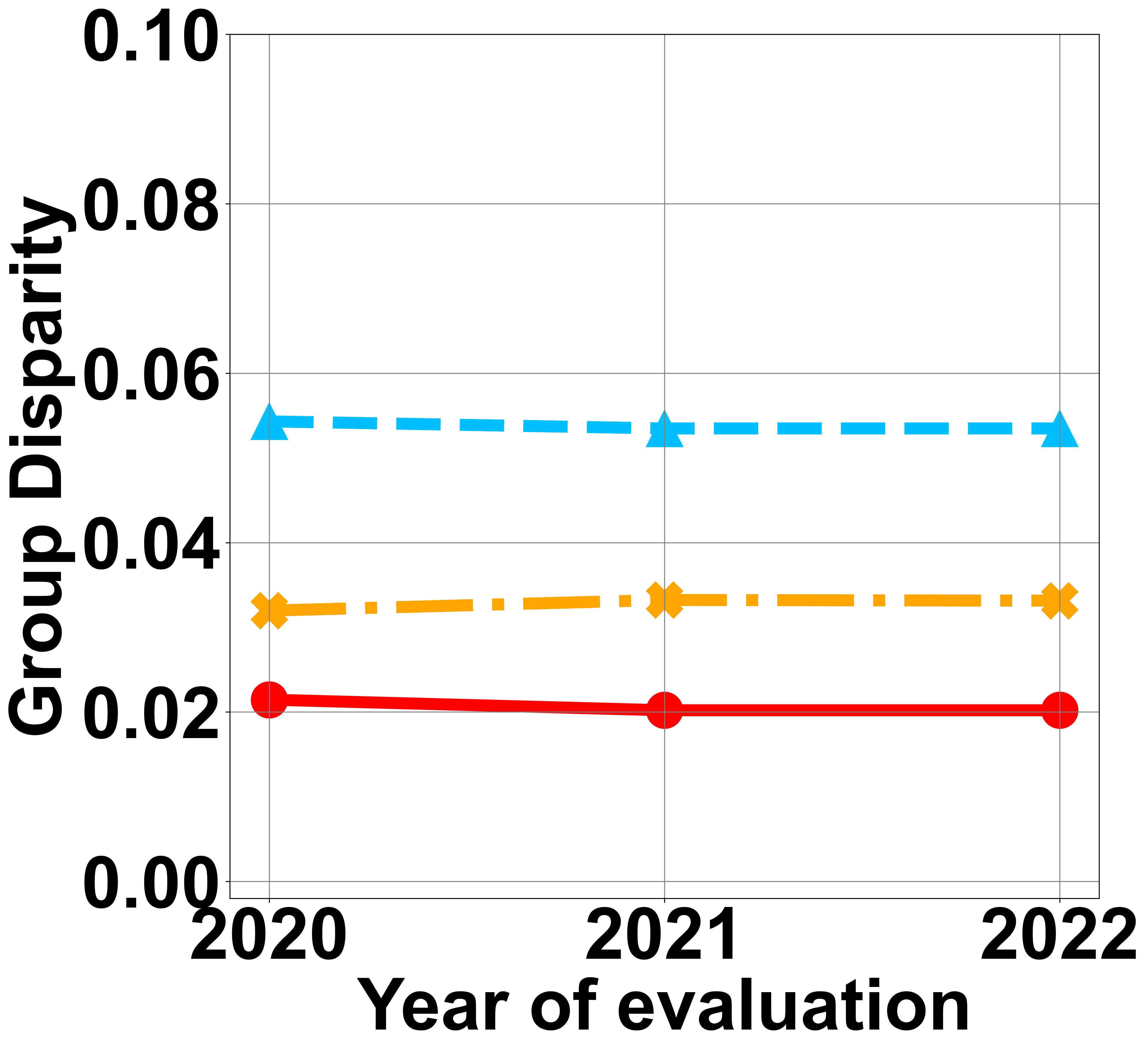}}
\end{subfigure}
\vspace{-0mm}
	\caption{{Additional accuracy and group disparity study. The two rows correspond to AMNIST and RAFDB respectively. Overall, there are  large accuracy differences and group disparity  for both cases. The group disparity on AMNIST is much larger than that on RAFDB, although thie former's accuracy is also higher. This further verifies that higher overall accuracy does not necessarily  lead to better fairness.   }}\label{fig:APIShift:additionalanalysis}
	\vspace{-0mm}
\end{figure}

We provide accuracy and group disparity study on two more datasets, AMNIST and RAFDB.
On AMNIST, we again use the accents of different speakers to group the datasets.
The accents in AMNIST cover ``German'', ``South Korean'', ``Spanish'', ``Madras'', ``Levant'', ``English'', ``Chinese'', ``Brasilian'', ``Italian'', ``Egyptian American'', ``South African'', ``Arabic'', ``Danish'', ``French'' and ``Tamil''.
On RAFDB, we use race (``Caucasian'', ``African-American'', and	 ``Asian'') to partition the dataset.
The results are shown in Figure \ref{fig:APIShift:additionalanalysis}.

Several interesting observation exist.
Overall, there are a large accuracy differences and group disparity  for both datasets. 
For instance, as shown in Figure \ref{fig:APIShift:additionalanalysis}(b), the group disparity of IBM can vary from 0.05 to 0.20. 
Noting the accuracy drop of IBM API during the same time period (2020-2021) is relatively smaller, one might infer that IBM's accuracy drop is due to worse ability to recognize certain accents. 
On the other hand, Microsoft API's overall accuracy on AMNIST seems to be stable (less than 0.3\% as shown in Figure \ref{fig:APIShift:additionalanalysis}(a)), but there is a significant change in its group disparity (larger than 3\% as shown in Figure \ref{fig:APIShift:additionalanalysis}(a)).
On RAFDB, the change over time is relatively smaller (Figure \ref{fig:APIShift:additionalanalysis}(c) and (d)). 
Yet, APIs with better accuracy exhibits lower group disparity. 
For example, Face++ API's accuracy is the lowest, and its group disparity is also the higest.
Thus, it still remains an interesting question to relate accuracy and group disparity changes due to API shifts.
How to determine which API or combination of APIs to use for different user data, budget constraints, accuracy and fairness targets is also enabled by \systemnameAPIBenchmark{} and open to the community.

\section{Additional Discussions}

\paragraph{Potential overfitting of commercial APIs on the publicly available datasets.} We suspect that the commercial APIs do not overfit the datasets we used for evaluations for three reasons. First, the terms of use for many of the datasets disallow commercial applications. For example, the RAFDB dataset is “available for non-commercial research purposes only” (see the webpage \url{http://www.whdeng.cn/RAF/model1.html}). Second, the performance of most evaluated APIs is well below that of typical overfitting, which is often more than 90\%. Third, we observed that several APIs’ performances dropped over time. For example, the EPixel API’s accuracy on the COCO dataset dropped from 47\% (Fall 2020) to 27\% (Spring 2022),  as shown in Figure \ref{fig:APIBenchmark:casestudy_coco} (a). This shows that it is still very interesting to compare commercial APIs over time on these datasets.

\paragraph{Licenses and  restrictions enforced by the ML APIs.}
The terms of use for most ML APIs (see, e.g., \url{https://cloud.google.com/terms} and \url{https://azure.microsoft.com/en-us/support/legal/}) require  no sublicensing  to a third party. However,  to the best of our knowledge, they do not prevent evaluating and analyzing those APIs' performance. In fact, evaluating and comparing the performance of different cloud services is not only desired by users but also encouraged by cloud providers. For example, Google Cloud provides its own performance measurement tool (\url{https://cloud.google.com/free/docs/measure-compare-performance#:~:text=Google\%20Cloud\%20Platform\%20provides\%20two,\%2Dto\%2Ddate\%20and\%20unbiased}). This is probably because a systematic study of the ML APIs can help the providers improve their services. For example, gender shade~\cite{pmlr-v81-buolamwini18a}, the seminal work on bias and stereotypes embedded in face detection APIs, has helped ML API providers improve their services and thus been appreciated by the industry.
We hope \systemnameAPIBenchmark{} enables better understanding of the commercial ML APIs and in turn helps API providers build better services too.

\paragraph{Maintenance and development plans for \systemnameAPIBenchmark{}.}
The maintenance and development plans consist of three main parts. 
First, we will continuously evaluate all ML APIs considered in the paper.
Currently the evaluation is planned to occur every 6 months. 
If significant performance changes are consistently observed every 6 months, the update frequency will be further increased, say, to every 3 months or every month.  
Second, we plan to enlarge the set of ML APIs, datasets, and tasks in \systemnameAPIBenchmark{}.
MLaaS is an increasingly growing industry, and new ML APIs are launched from time to time.
Thus, we plan to add the evaluation of the emerging ML APIs every 6 months.
It is also important to include more representative and diverse datasets and document how quality of the datasets affects ML APIs' performance.  
For example, for image classification, ML APIs' robustness to the image resolution and natural noises (such as rain and snow) can largely influence practitioners' choices.
Last but not least, the usefulness of a database is determined by our community. 
Thus, we plan to implement an interactive 
feedback system on our website to collect opinions from our community. 
This helps, for example, solicit preference of which datasets, ML APIs, and tasks to include in \systemnameAPIBenchmark{}.

\begin{figure} \centering
\begin{subfigure}[PASCAL]{\label{fig:sarsity_a}\includegraphics[width=0.32\linewidth]{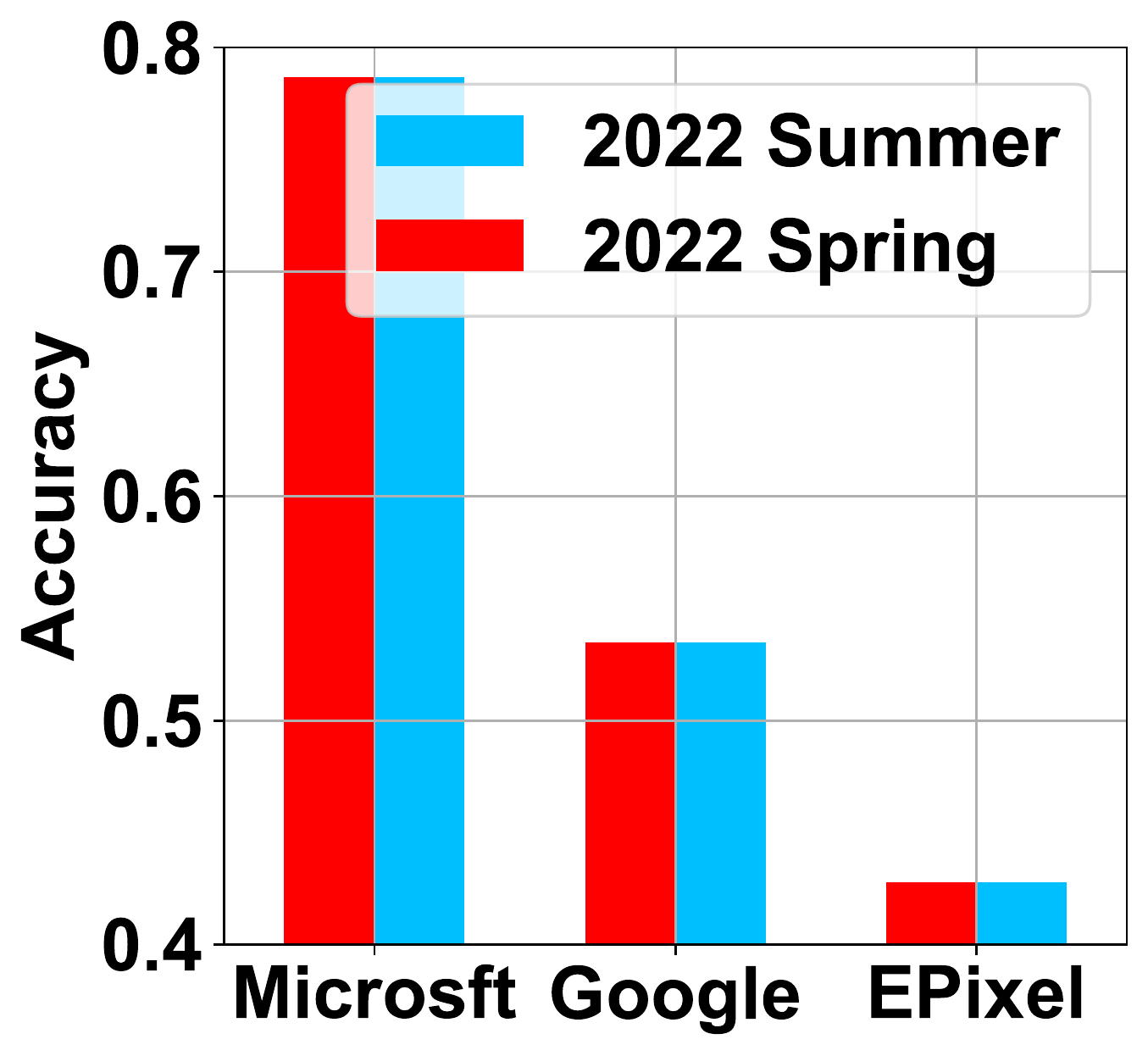}}
\end{subfigure}
\begin{subfigure}[MIR]{\label{fig:sparsity_b}\includegraphics[width=0.32\linewidth]{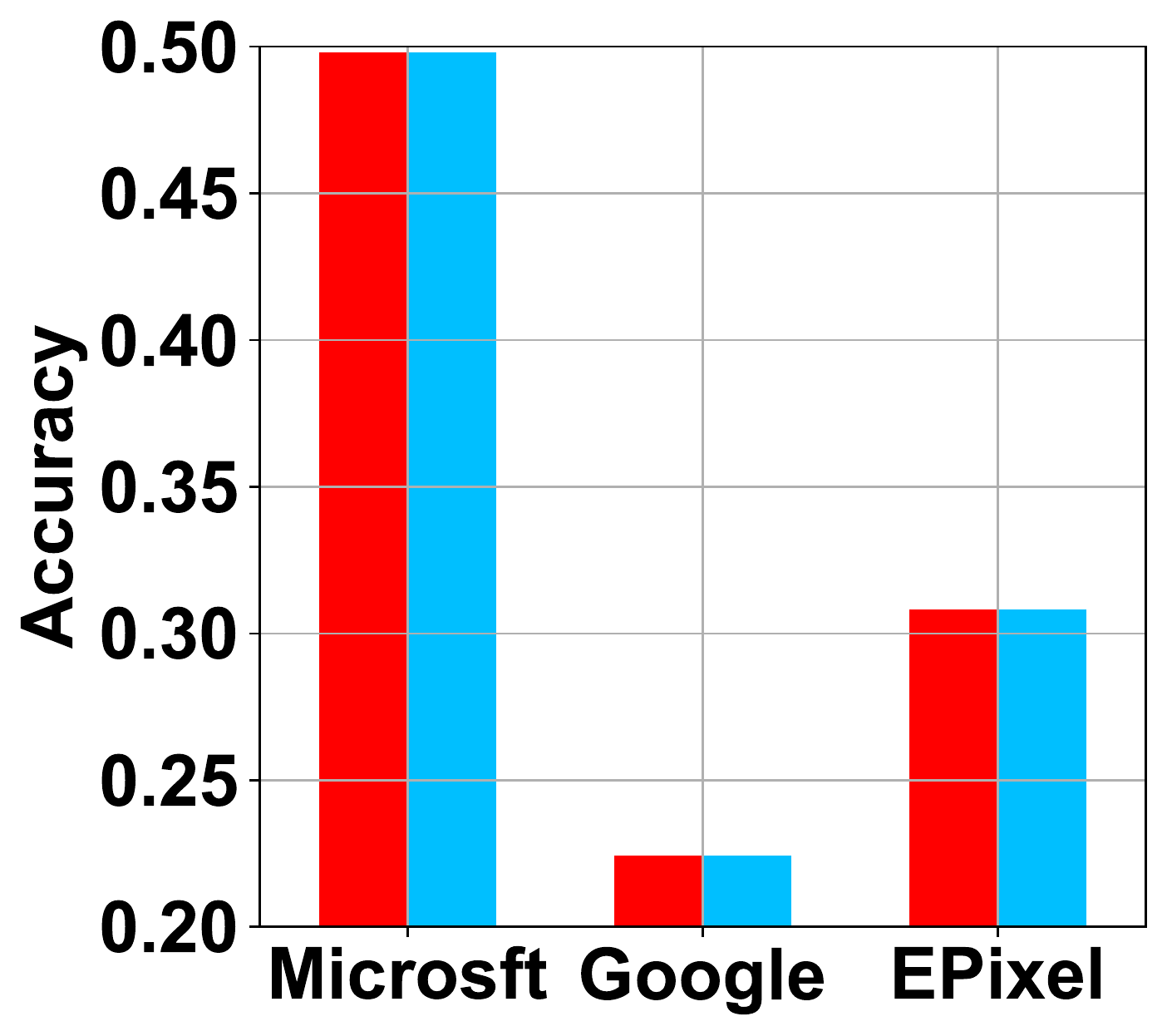}}
\end{subfigure}
\begin{subfigure}[COCO]{\label{fig:sparsity_c}\includegraphics[width=0.32\linewidth]{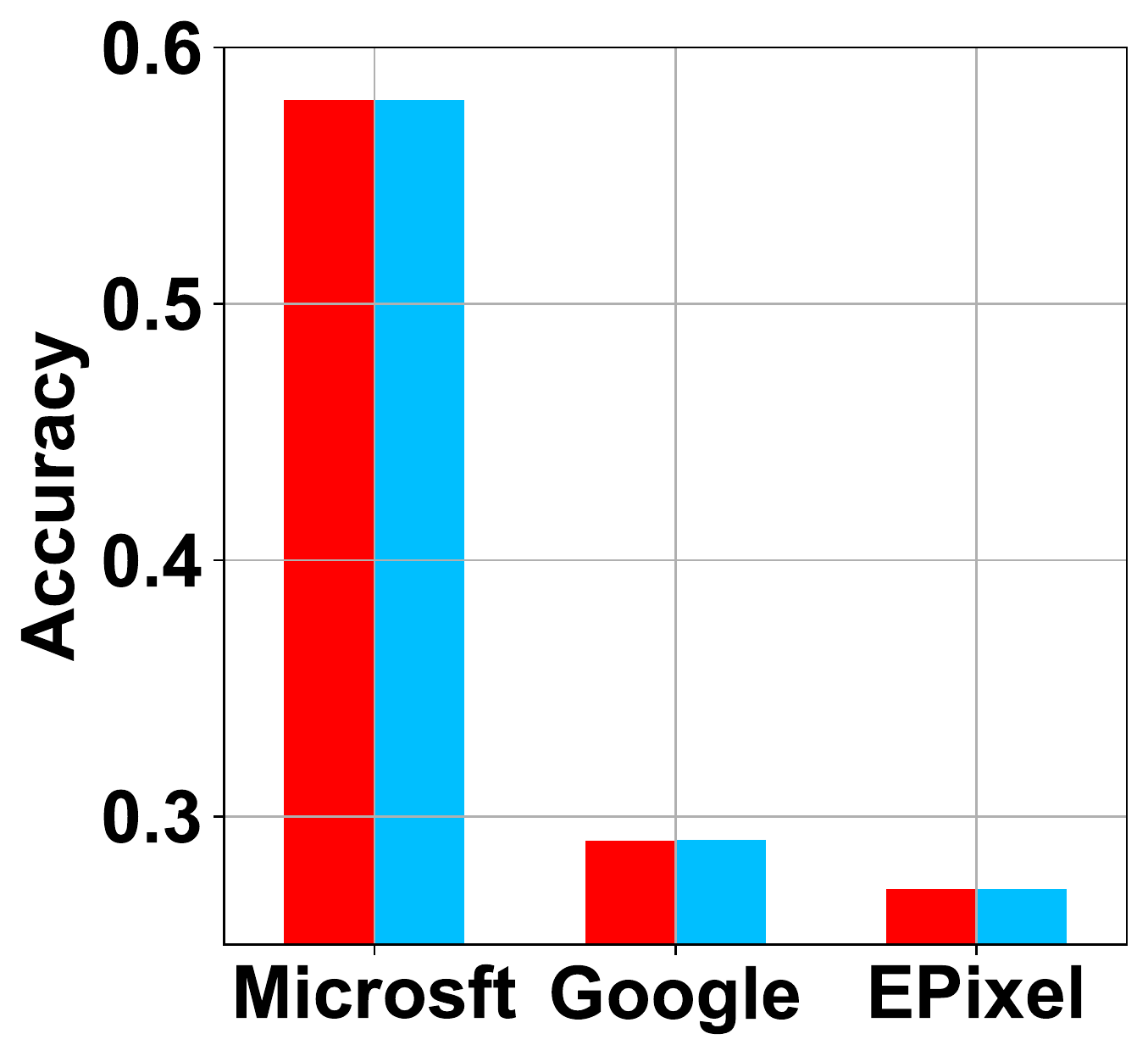}}
\end{subfigure}

\begin{subfigure}[CONLL]{\label{fig:sparsity_b}\includegraphics[width=0.32\linewidth]{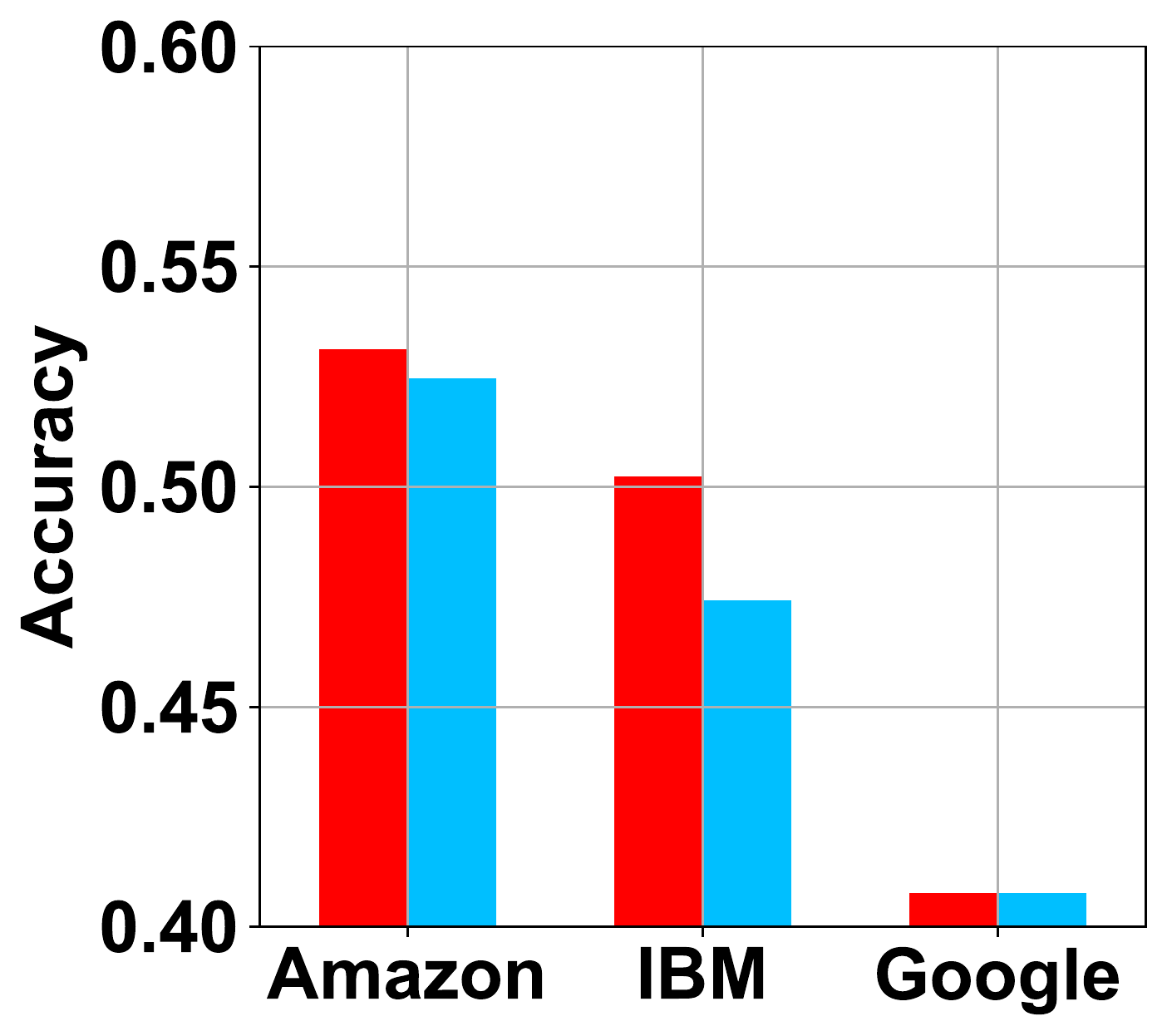}}
\end{subfigure}
\begin{subfigure}[GMB]{\label{fig:sarsity_a}\includegraphics[width=0.32\linewidth]{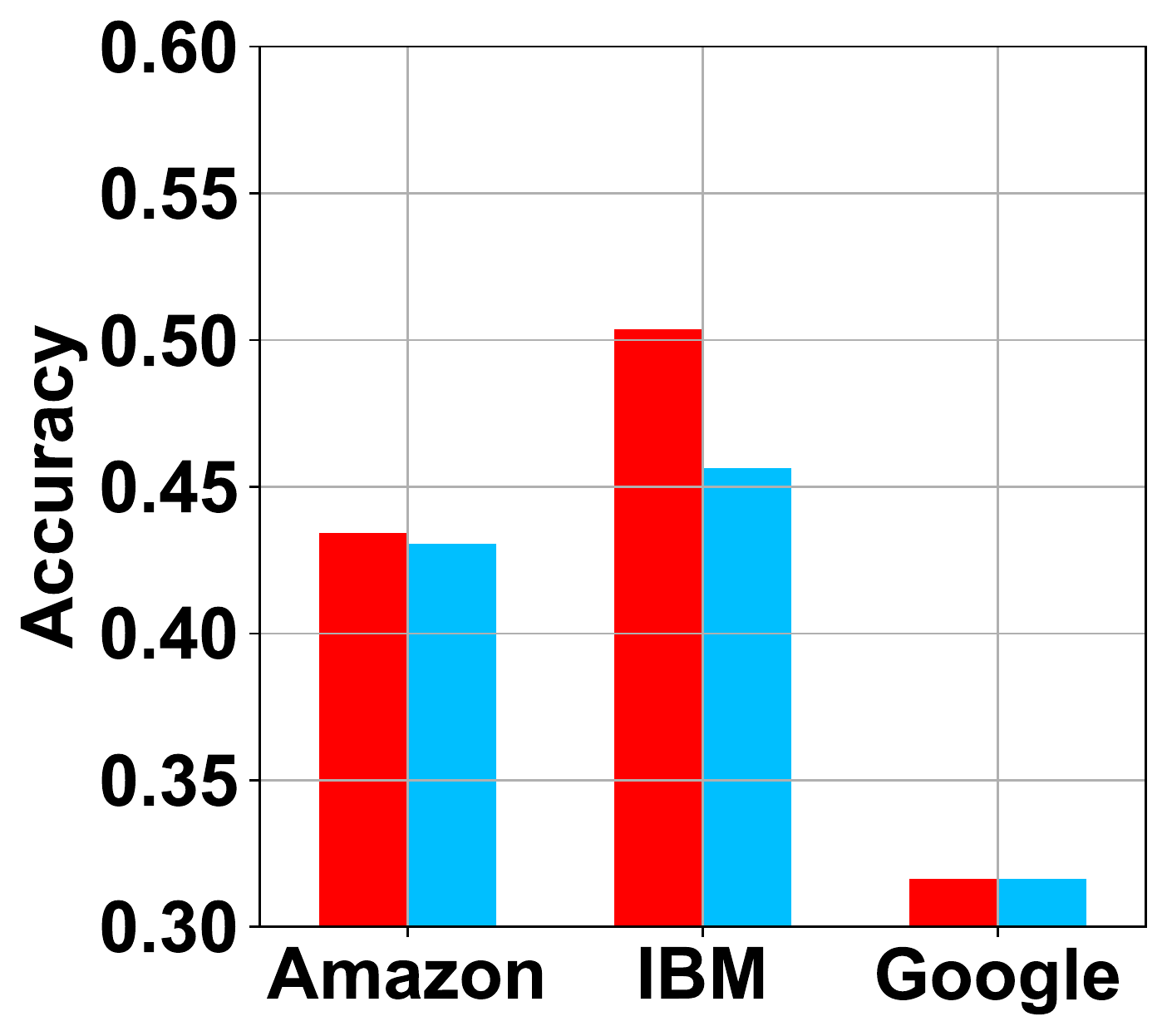}}
\end{subfigure}
\begin{subfigure}[ZHNER]{\label{fig:sparsity_c}\includegraphics[width=0.32\linewidth]{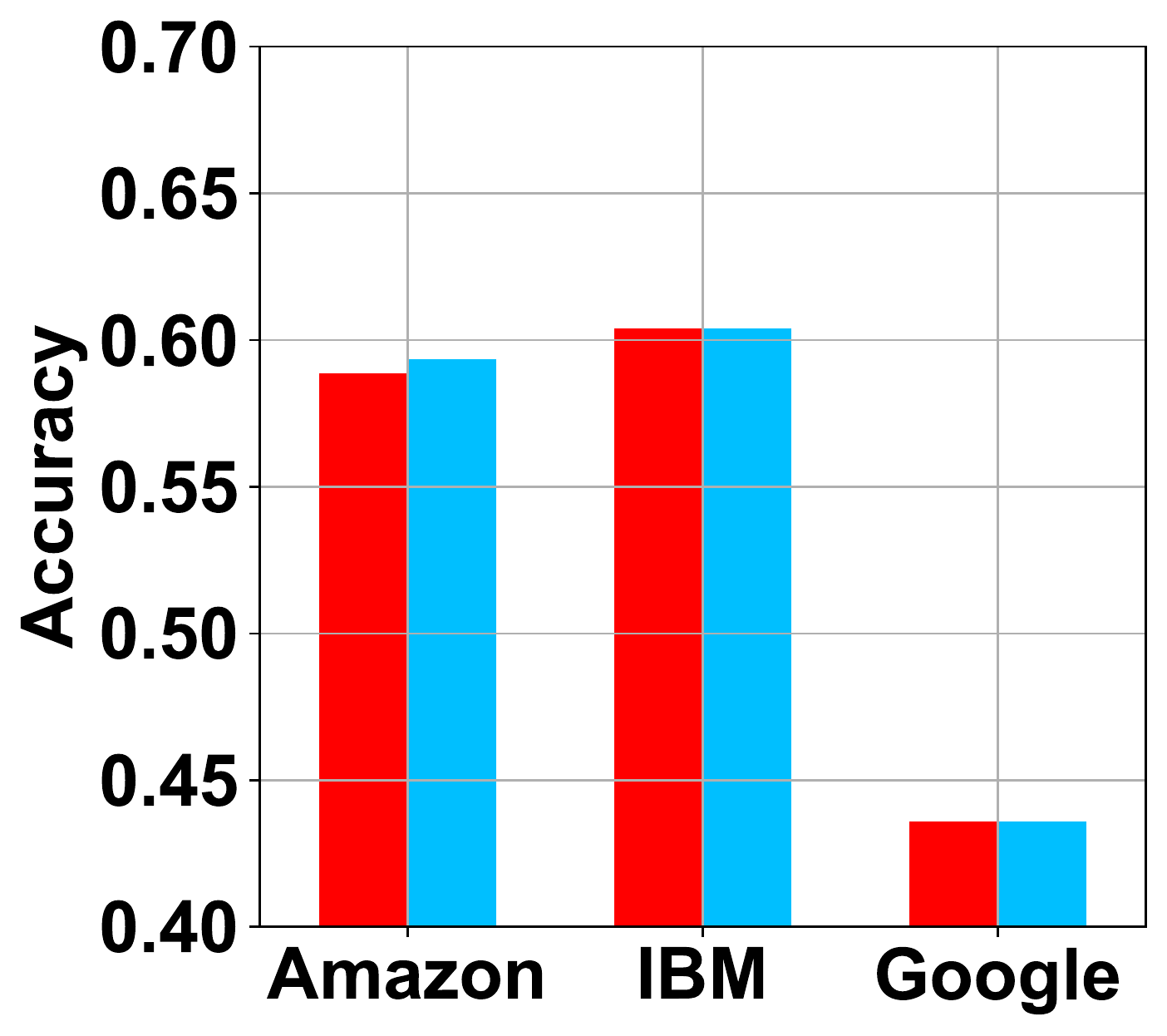}}
\end{subfigure}

\begin{subfigure}[MTWI]{\label{fig:sarsity_a}\includegraphics[width=0.32\linewidth]{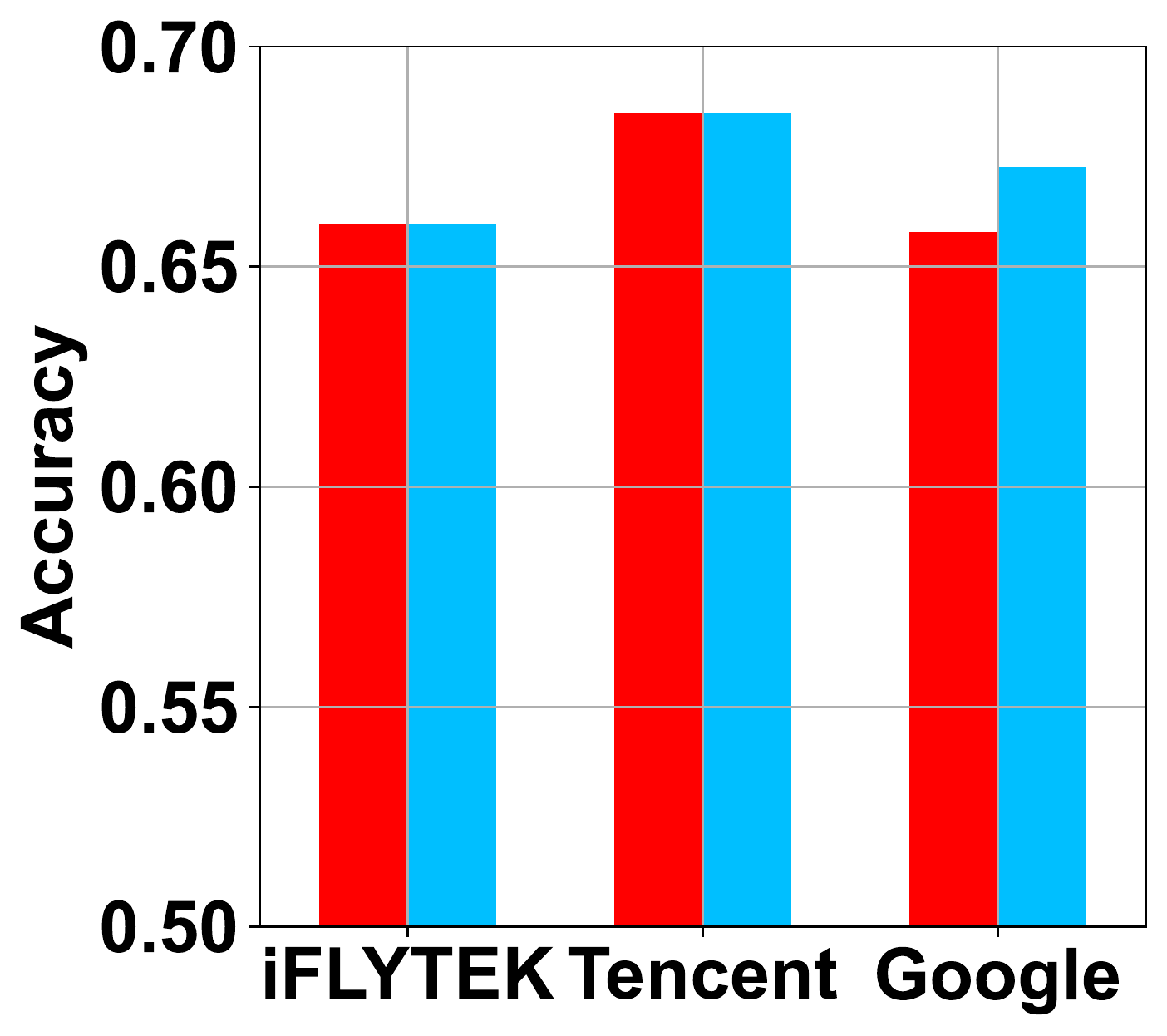}}
\end{subfigure}
\begin{subfigure}[ReCTS]{\label{fig:sparsity_b}\includegraphics[width=0.32\linewidth]{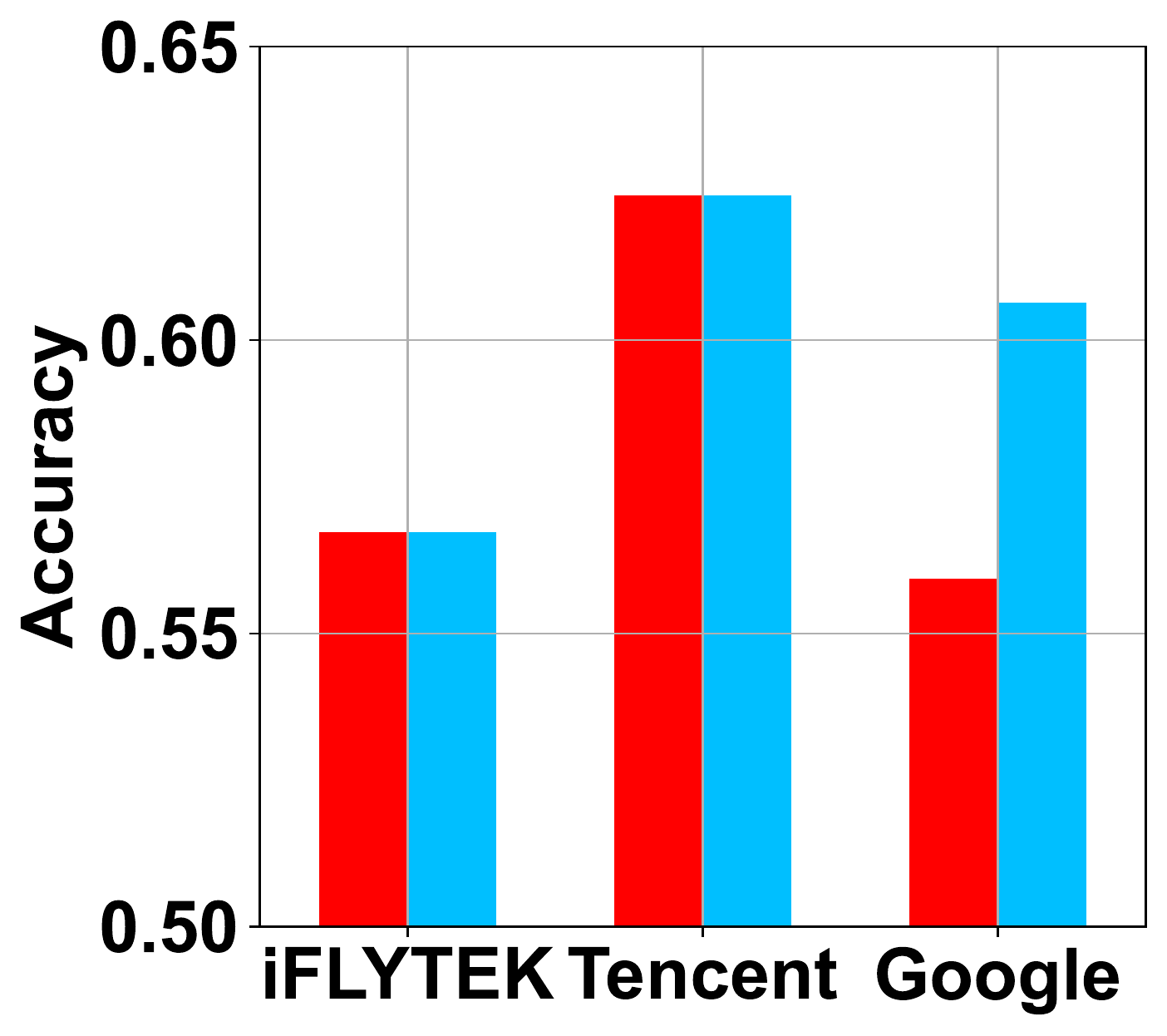}}
\end{subfigure}
\begin{subfigure}[LSVT]{\label{fig:sparsity_c}\includegraphics[width=0.32\linewidth]{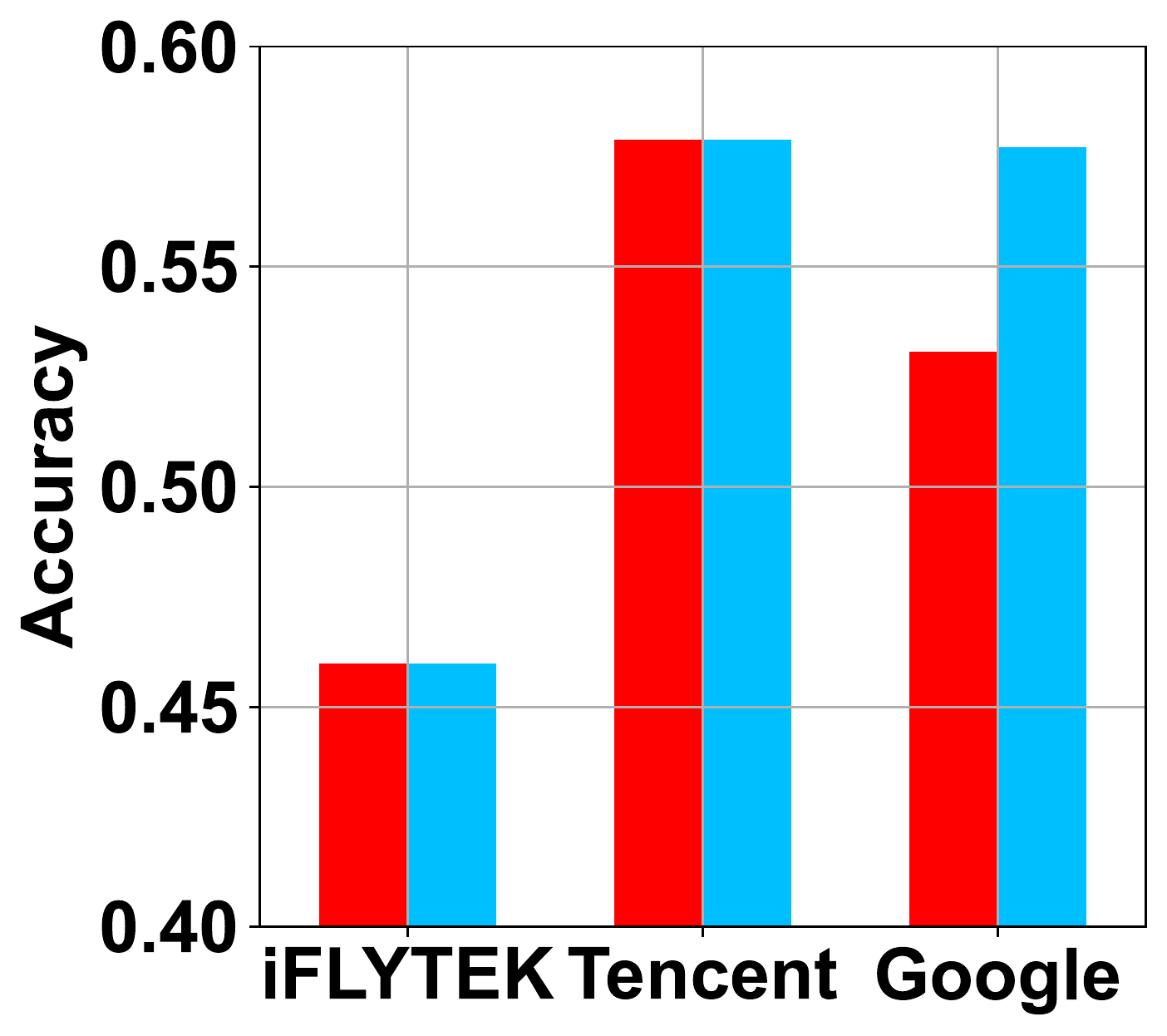}}
\end{subfigure}

\vspace{-0mm}
	\caption{{Accuracy changes of structured prediction APIs within 6 months (2022 Spring -- 2022 Summer). The first, second, and third row corresponds to multi-label image classifications, scene text recognition, and named entity recognition. Overall, we observe accuracy shifts of  several APIs. 
	For example, IBM named entity API's performance dropped on CONLL and GMB, while the accuracy of Google scene text API increased on MTWI and ReCTS. }}\label{fig:APIBenchmark:6month_acc}
\end{figure}

\begin{figure} \centering
\begin{subfigure}[PASCAL]{\label{fig:sarsity_a}\includegraphics[width=0.32\linewidth]{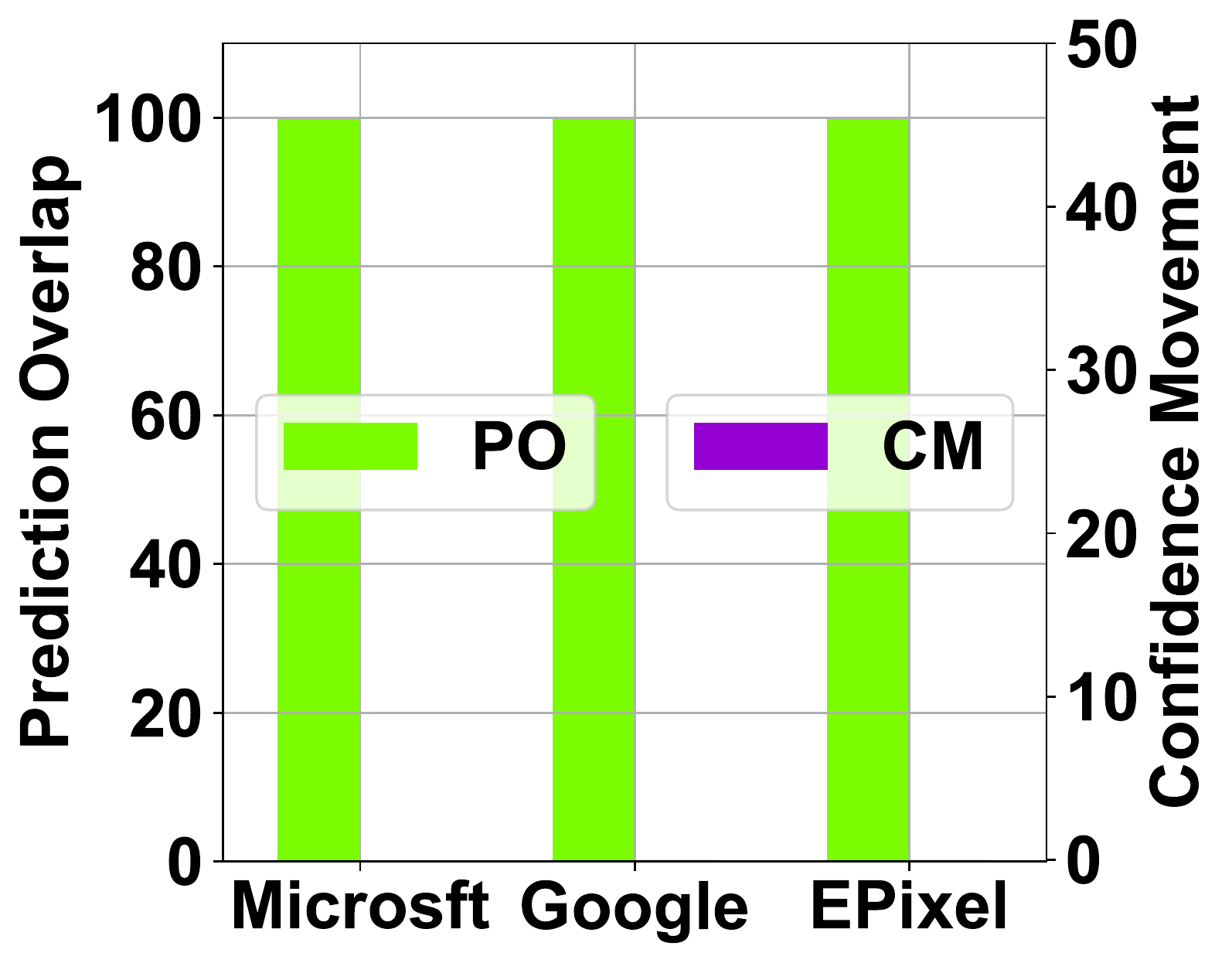}}
\end{subfigure}
\begin{subfigure}[MIR]{\label{fig:sparsity_b}\includegraphics[width=0.32\linewidth]{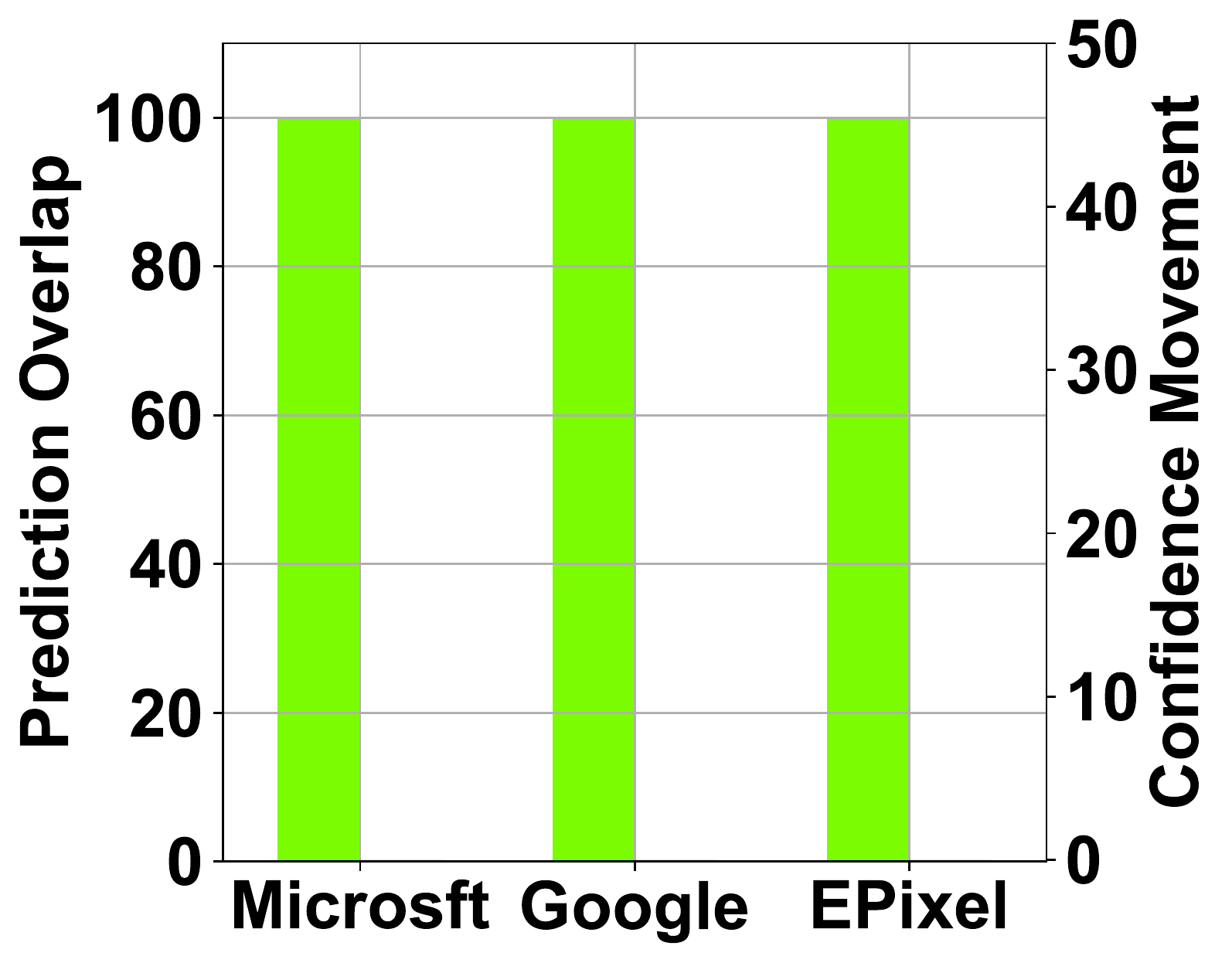}}
\end{subfigure}
\begin{subfigure}[COCO]{\label{fig:sparsity_c}\includegraphics[width=0.32\linewidth]{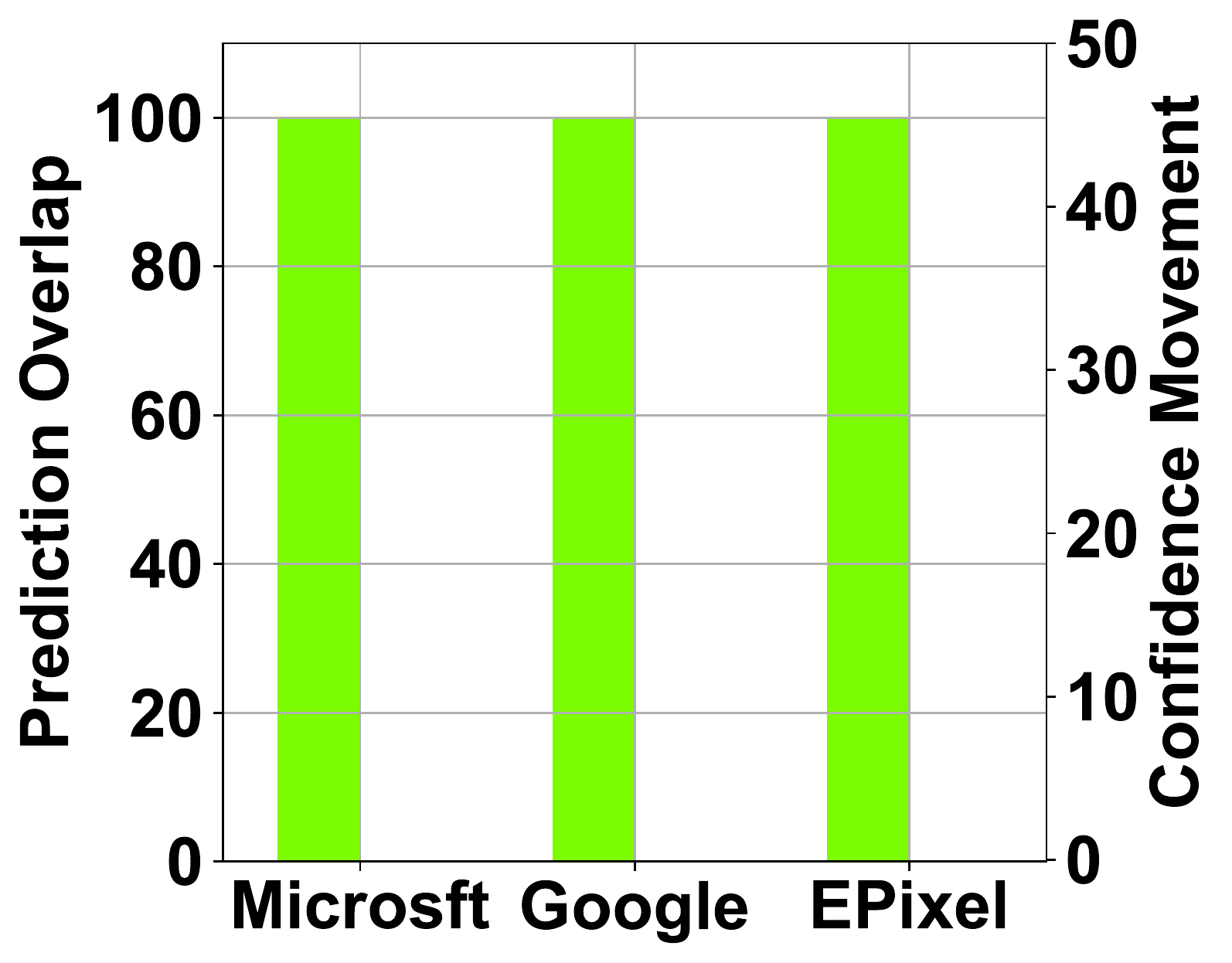}}
\end{subfigure}

\begin{subfigure}[CONLL]{\label{fig:sparsity_b}\includegraphics[width=0.32\linewidth]{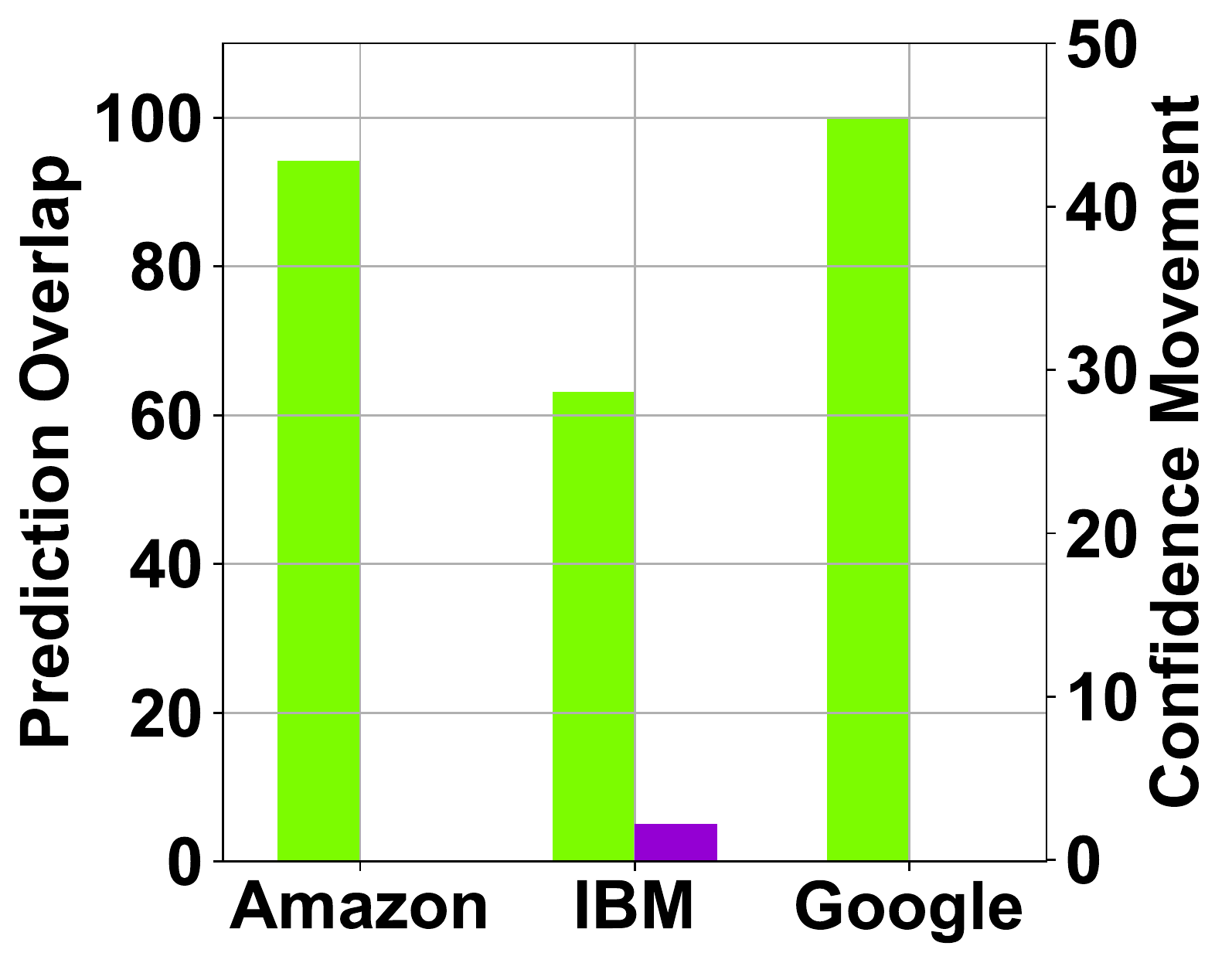}}
\end{subfigure}
\begin{subfigure}[GMB]{\label{fig:sarsity_a}\includegraphics[width=0.32\linewidth]{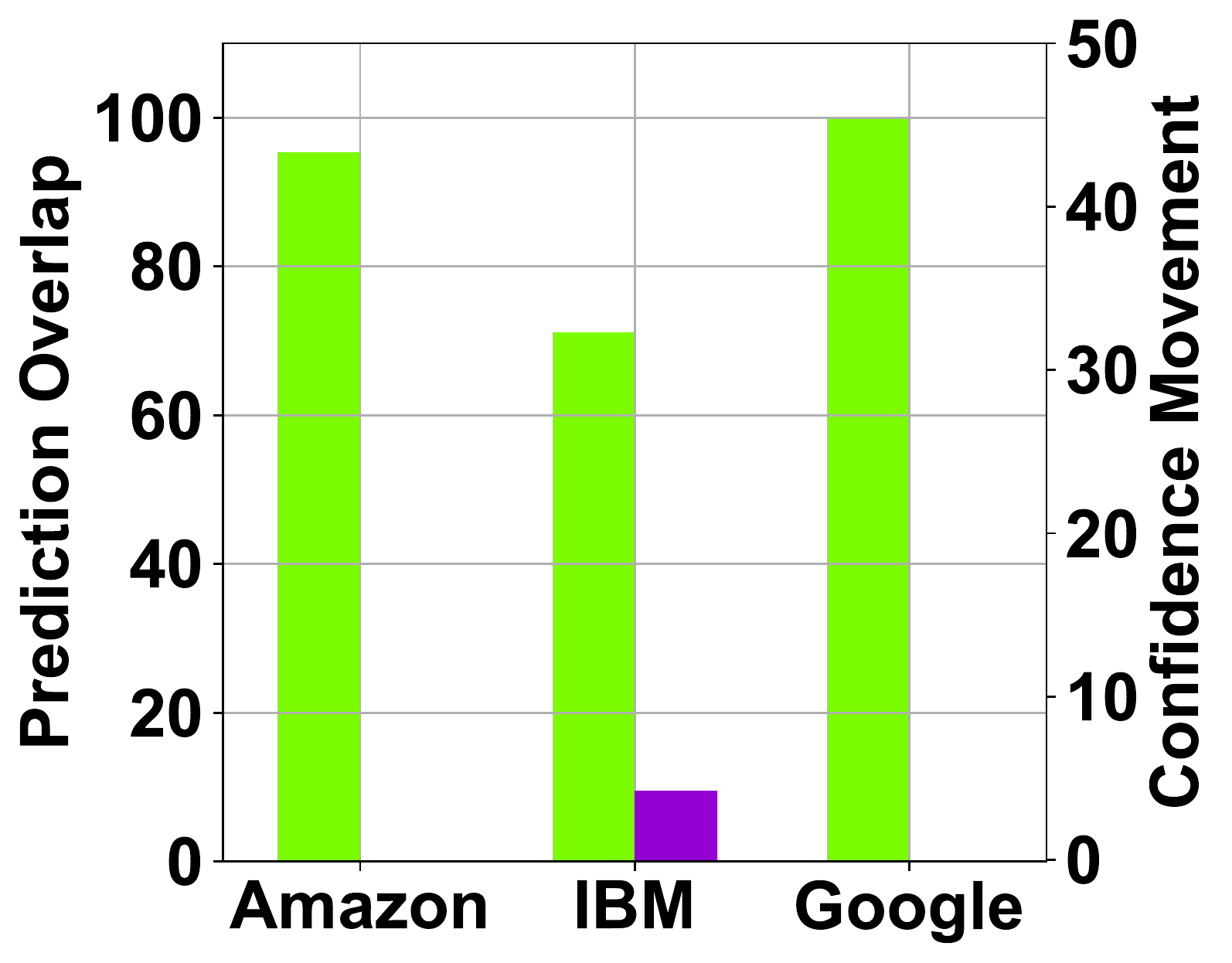}}
\end{subfigure}
\begin{subfigure}[ZHNER]{\label{fig:sparsity_c}\includegraphics[width=0.32\linewidth]{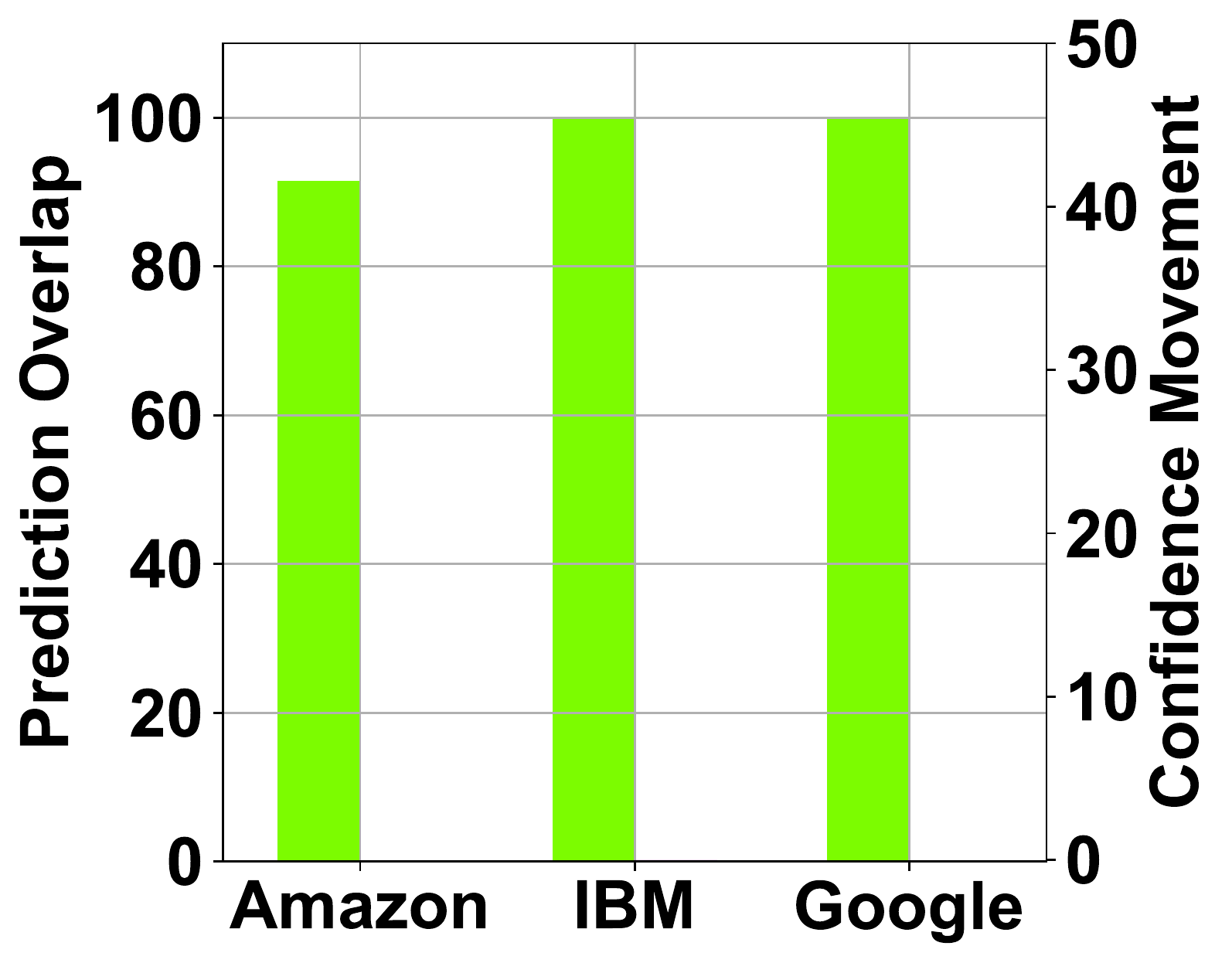}}
\end{subfigure}

\begin{subfigure}[MTWI]{\label{fig:sarsity_a}\includegraphics[width=0.32\linewidth]{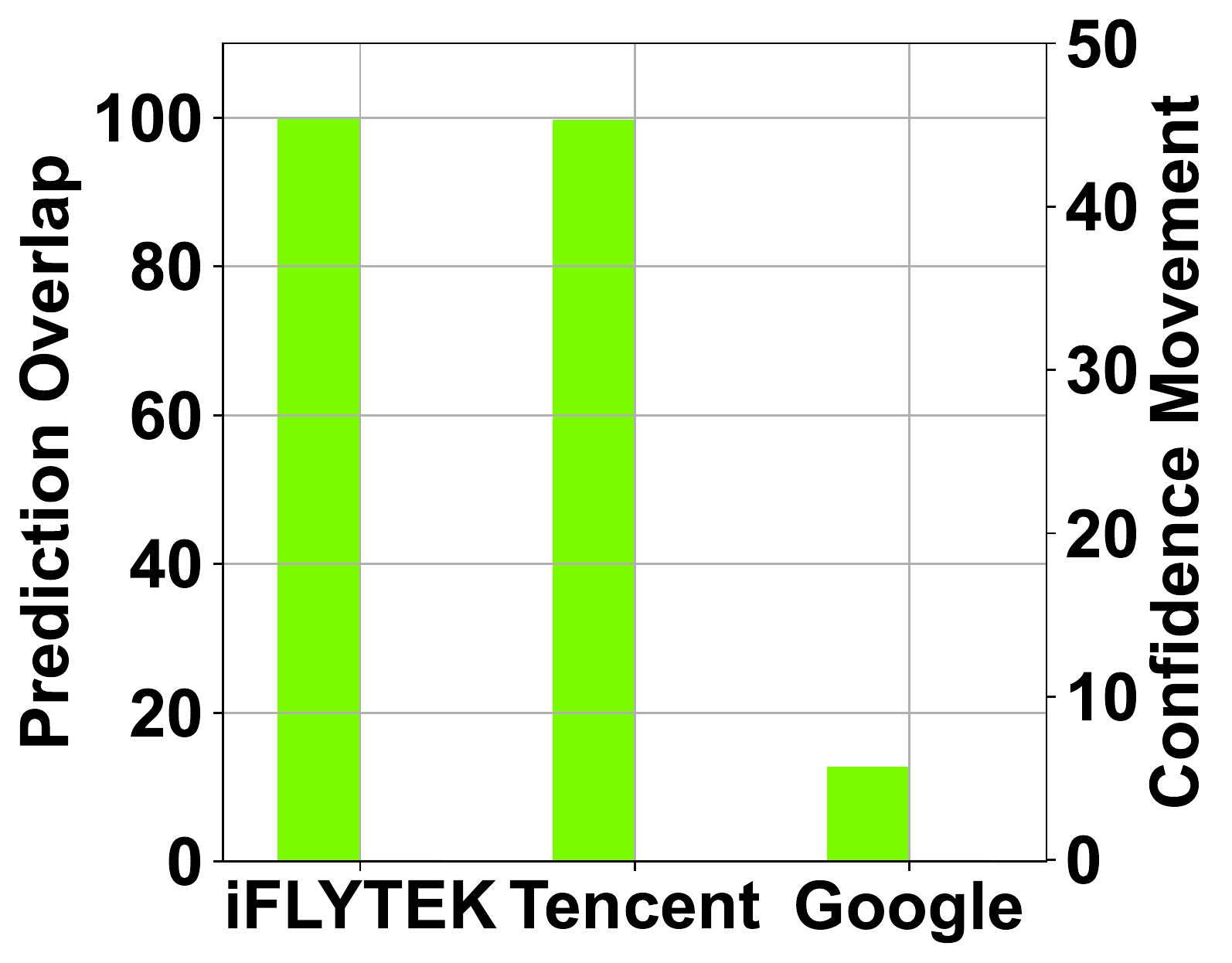}}
\end{subfigure}
\begin{subfigure}[ReCTS]{\label{fig:sparsity_b}\includegraphics[width=0.32\linewidth]{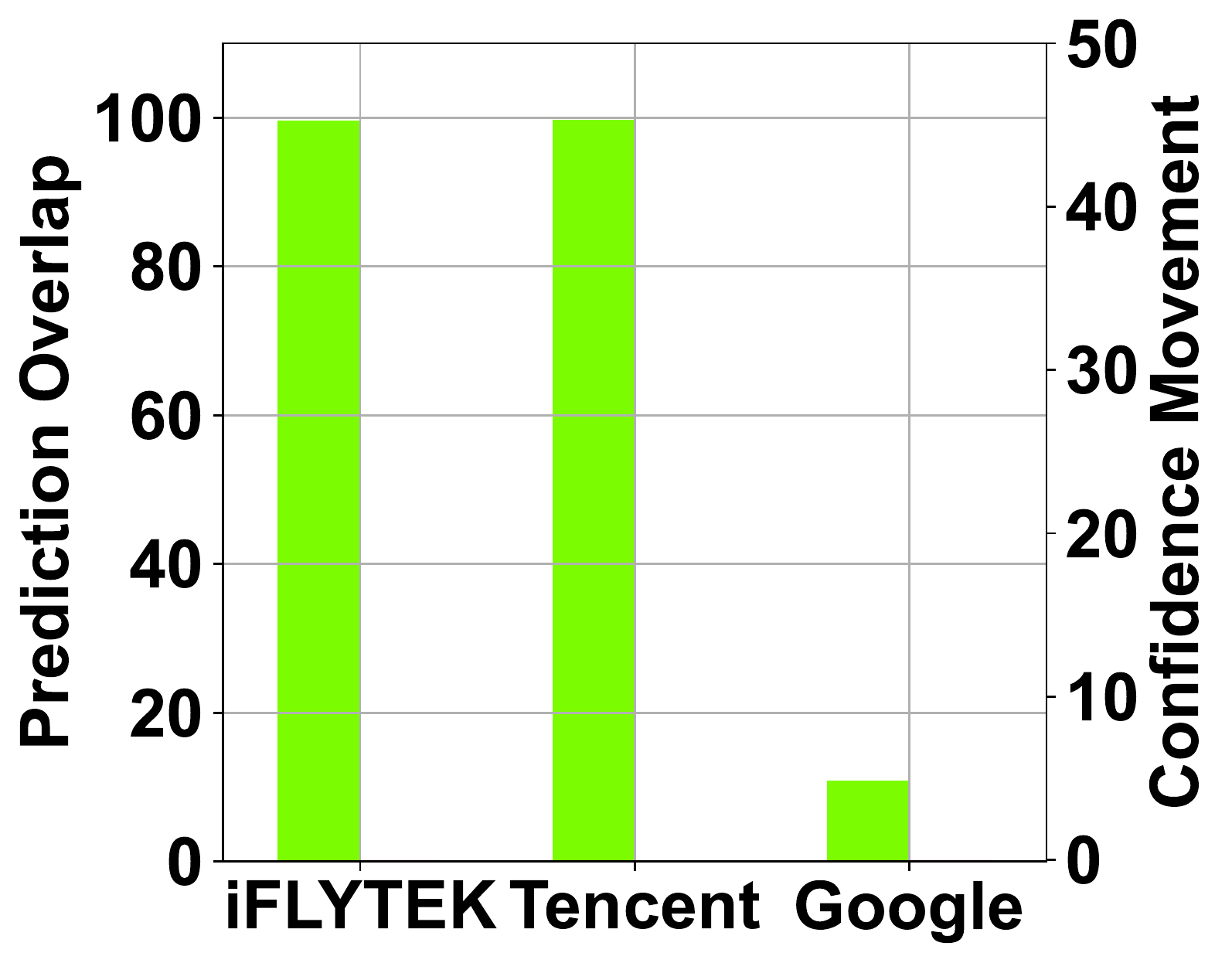}}
\end{subfigure}
\begin{subfigure}[LSVT]{\label{fig:sparsity_c}\includegraphics[width=0.32\linewidth]{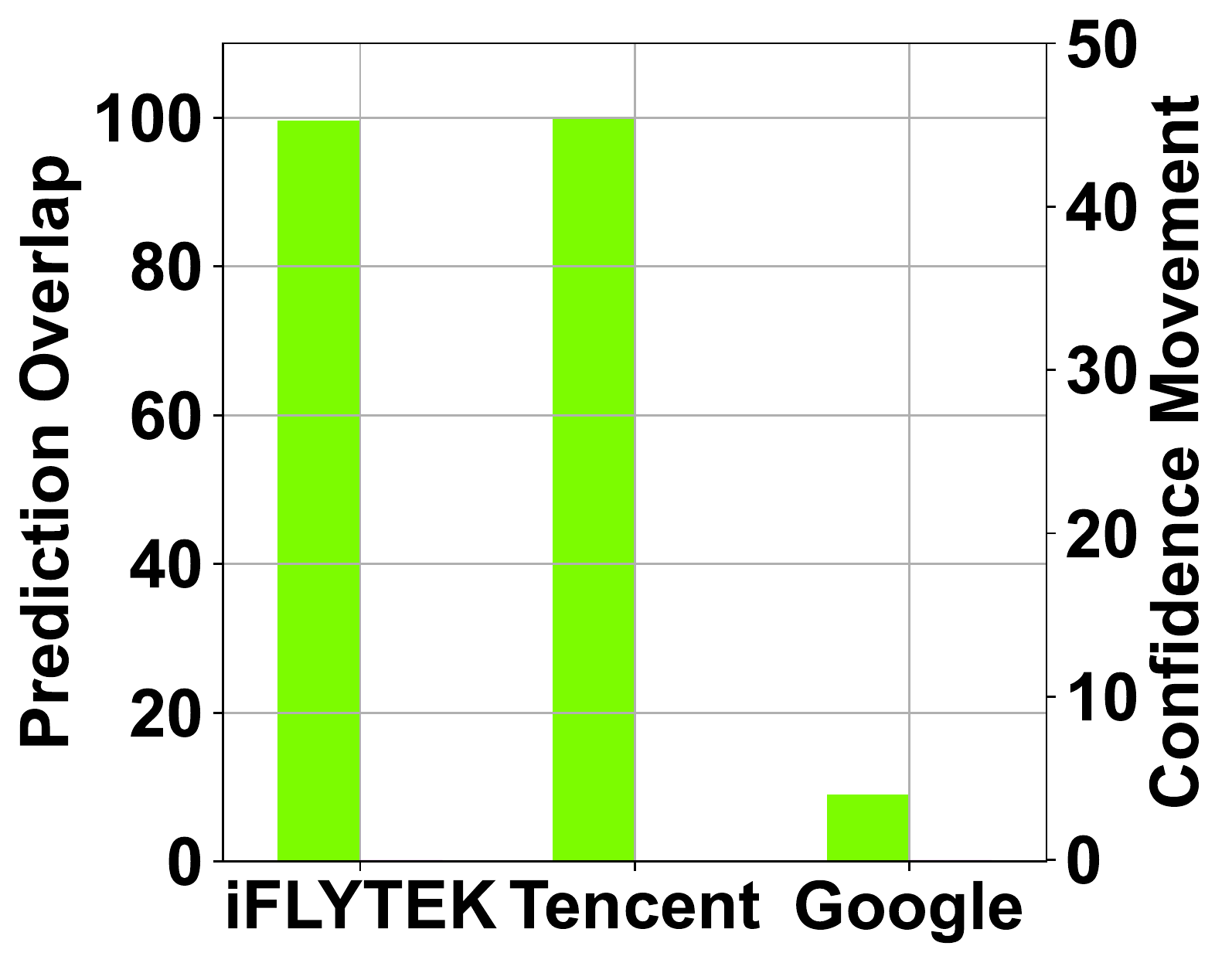}}
\end{subfigure}

\vspace{-0mm}
	\caption{{Prediction overlap and confidence movement  of structured prediction APIs within 6 months (2022 Spring -- 2022 Summer). The first, second, and third row correspond to multi-label image classifications, scene text recognition, and named entity recognition. Overall, most prediction shifts occurred for scene text and named entity recognition. 
	There was little prediction change of multi-label image classification APIs. 
	The confidence scores of the IBM API increased on CONLL and GMB and remained almost the same for the other APIs.  }}\label{fig:APIBenchmark:6month_confidenceandoverlap}
\end{figure}

As a first step, we have collected the predictions from ML APIs for all structured tasks, including multi-label image classification, scene text recognition, and named entity recognition,  in August 2022. 
The accuracy changes and prediction overlap as well as confidence movement compared to the prediction collected 6 months ago (February or March 2022) are shown in 
Figure \ref{fig:APIBenchmark:6month_acc} and  Figure \ref{fig:APIBenchmark:6month_confidenceandoverlap}, respectively. Overall, we observe   
accuracy shifts of  several APIs.
For example, the accuracy of the IBM API for named entity recognition on the GMB dataset dropped from 50\% (March 2022) to 45\% (August 2022), as shown in Figure \ref{fig:APIBenchmark:6month_acc} (e). The performance of the Google scene text recognition API was 60\% in the ReCTS dataset in August 2022, which was 4\% higher than that in March 2022 as shown in Figure \ref{fig:APIBenchmark:6month_acc} (h). In fact, prediction changes of the Google API occurred on more than 80\% of images in ReCTS as well as MTWI and ReCTS, as shown in Figure \ref{fig:APIBenchmark:6month_confidenceandoverlap} (g), (h) and (i). There was little prediction changes of multi-label image classification APIs. 
The confidence scores of the IBM API increased on CONLL (by 4\%) and GMB (by8\%) and remained almost the same for the other APIs, as shown in Figure \ref{fig:APIBenchmark:6month_confidenceandoverlap} (d) and (e). 
Overall, this analysis suggests that significant changes can happen within  six months and thus  frequent updates of the database is needed. 

\paragraph{Choices of Datasets and ML APIs. }
\eat{The datasets and ML APIs were chosen based on (i) popularity within the ML community, (ii) easy access from the internet, and (iii) diverse representations.
For example, while popular ML APIs from Google and Microsoft were evaluated, our database also includes API predictions from less-representative companies such as EPixel. 
It is worthy noting that the terms of uses for many datasets requires no commerical usages. 
In addition, we have also observed that many commercial APIs's performances dropped over time. 
Thus, little dataset overfitting should have occurred within those  commercial APIs.
}
We chose existing datasets for a few reasons. First, the ML community is familiar with the datasets and they are relatively well annotated and evaluated. Second, those datasets can be easily assessed on the internet. Third, those datasets covered a diverse range of real-world scenarios (for example, the COCO dataset included objects in outdoor/indoor environments, at a small/large scale, and with different brightness). In fact, based on our conversation with many practitioners, there is a large interest in understanding commercial APIs’ performance on those datasets. Thus it is a good starting point to evaluate ML APIs on those popular datasets. 
 
Similarly, the selection criteria for ML APIs are (i) popularity, (ii) easy access for users, and (iii) representation of diverse companies. Based on our discussion with practitioners, Google APIs are widely used and easily accessible and hence included in our database. ML APIs from domain-specific companies such as EPixel, Face++, and iFLYTEK were also included to increase the representativeness of our database.

\paragraph{Responsible usage of facial emotion datasets.} According to the original documents of the facial emotion datasets (FER+~\cite{dataset_FERP_BarsoumZCZ16}, RAFDB~\cite{Dataset_FAFDB_li2017reliable},
EXPW~\cite{Dataset_EXPW_SOCIALRELATION_ICCV2015}, and AFNET~\cite{Dataset_AFFECTNET_MollahosseiniHM19}), all the face images in these four facial emotion datasets were collected via querying search engines (e.g., Google, Bing, and Yahoo!) with certain keywords (e.g., happy faces). While the images are publicly retrievable from search engines, we did not find clear documentation of the individual consent process for these datasets. We recognize that facial photos are sensitive data, and will remove photos from HAPI upon request. Moreover, photos curated online may not fully represent the general public, and emotion annotations can be subjective and noisy. Therefore, analysis of these datasets should be interpreted with care. For example, the fact that an API’s performance on some of these datasets changes over time is important to know, while the absolute performance across different datasets may not be directly comparable. We will continue to work with the machine learning community to expand HAPI to include high-quality benchmark datasets.

\paragraph{Additional outputs from ML APIs.}
Several APIs generate information beyond confidence scores and predicted labels. 
For instance, for  multi-label image classification,  Microsoft vision API provides the bounding boxes for all detected objects. 
Given a text paragraph, Google sentiment analysis API
returns not only a predicted attitude label with a confidence score, but also a magnitude score indicating how significant the detected attitude is. 
\systemnameAPIBenchmark{}  allows users to query the raw outputs including the above information, too.

\paragraph{Relations to model stealing attacks and defenses.} Model stealing attacks~\cite{ModelStealAttack_orekondy2019knockoff,ModelStealingAttack_tramer2016stealing} and defenses~\cite{_ModelStealingDefense_juuti2019prada, ModelStealingDefense_orekondy2019prediction} have raised large attentions in both security and ML communities. 
\systemnameAPIBenchmark{} provides a large set of predictions from real-world ML APIs to study model stealing attacks and defenses.
An interesting next step, for example, 
is to benchmark  different model stealing attacks on \systemnameAPIBenchmark{}. 
It is also interesting to study if applying model inversion attacks~\cite{ModelInversionAttack_fredrikson2015model} on the stolen model can steal the training datasets of commercial ML APIs.

\paragraph{Strength and weakness of the evaluated datasets.}

Recall that our dataset selection criteria are (i) popularity, (ii) easy access, and (iii) diversity. Now we provide more details about how the selected datasets meet the criteria and what limitations remain. 
We start by the speech command recognition datasets. They are all widely studied by the speech command recognition community (for example, CMD~\cite{Dataset_Speech_GoogleCommand} has been cited more than 700 times since published in 2018) and are easily accessible on the internet. They contain a diverse range of commands: DIGIT and AMNIST mainly obtain spoken digits, while CMD and FLUENT contain more complicated commands such as “turn on the light in the kitchen”. Their varying sampling rates also cover different application scenarios. In addition, speaker information is also provided, enabling fairness study. A potential limitation is that all those datasets are clean, i.e., there is almost no environmental noise in the utterances. Evaluating ML APIs’ robustness to such noise is an interesting next step.
Similarly, the datasets for SA and FER are also widely used. For example, the IMDB~\cite{Dataset_SEntiment_IMDB_ACL_HLT2011} dataset has been cited more than 3,000 times, and the citation of  RAFDB~\cite{Dataset_FAFDB_li2017reliable} is above 700.
Besides easy access on the internet, they also represent diverse data distribution: YELP, IMDB, WAIMAI cover user reviews for restaurants, movies, and delivery services, while feedback for items from various category is included in SHOP.
Limitations include (i) that only English and Chinese texts are included, and (ii) that most text paragraphs are short. 
FER+ contains gray and low-resolution images, while RAFDB, EXPW and AFNET consist of colored images with high resolutions.
One limitation is that only few images contain more than one person.

Similarly, the datasets used for the structured prediction tasks are also widely used and easily accessible. 
For example, PASCAL~\cite{Dataset_Pascal_2015} and COCO~\cite{Dataset_COCO_2014} are perhaps the most widely studied datasets for object recognition. 
MTWI~\cite{Dataset_MTWI_2018}, ReCTS~\cite{Dataset_ReCTS_2019}, and LSVT~\cite{Dataset_LSVT_2019} are one of the largest scene text recogntion datasets and were used for competitions in
International Conference on Document Analysis and Recognition (ICDAR), one of the flagship conferences  on document analysis.
CONLL~\cite{Dataset_CONLL_2003}  and GMB~\cite{Dataset_GMB_2013}
are widely studied for named entity recognition in English, while the GitHub repository hoding the Chinese named entity recongition dataset,
ZHNER~\cite{ZHNER_dataset_github}, has received almost two thousand stars. 
Besides easy access, they also cover different scenarios. 
For example, most images are low-resolution in PASCAL but high-resolution in COCO and MIR.
MTWI contains mostly advertising images, while most images are photos taken on sign boards in ReCTS and on street view in LSVT. 
A natural way to extend the diversity of the datasets is to evaluate vision APIs on images with large number of labels (e.g., larger than 1000).
It is also interesting to study how ML APIs perform on multilingual scene text images.
Multilingual text datasets with domain specifications are also useful to understand named entity recognition APIs. 
Continuously identifying and evaluating ML APIs on  more diverse datasets is part of our future plans.

\paragraph{Support of AI ethics.} \systemnameAPIBenchmark{} enables the study of AI ethics on a range of  commercial systems targeting various tasks. 
For example, predictions of vision APIs on human objects can be used to study the biases and stereotypes on sensitive features including races, genders, and ages.
The evaluation of speech APIs opens the door for understanding and comparing how accents and nationality of the speakers affect different ML APIs' performance. 
Besides understanding the real-world APIs' ethic issues, how to efficiently detect and estimate those issues can also be explored on top of   \systemnameAPIBenchmark{}.
For example, one may use the heterogeneity of the predicted labels between different population groups to detect an API’s biases.
In addition, \systemnameAPIBenchmark{} offers an opportunity to explore whether and how the biases and stereotypes can be mitigated by adaptively selecting which API to use. 
In a nutshell,  \systemnameAPIBenchmark{} supports various studies of trustworthy AI  on a range of  commercial APIs.

\section{Datasheet}

This section includes a ``datasheet" for the dataset, following the outline proposed by  \cite{gebru2021datasheets}. 

\subsection{Motivation}
\textbf{For what purpose was the dataset created?} \systemnameAPIBenchmark{} was created to enable research on ML APIs. This includes but is not limited to, for example, determining which API or combination of APIs to use for different user data or applications as well as budget constraints, estimating how much performance has changed due to API shifts, and  explaining  the performance gap due to ML API  shifts.

\textbf{Who created the dataset?} The dataset was created in the Zou Group at Stanford University. 

\textbf{Who funded the creation of the dataset?} This project is supported in part by NSF CCF 1763191, NSF CAREER AWARD 1651570 and NSF CAREER AWARD 1942926.

\subsection{Composition}

\textbf{What do the instances that comprise the dataset represent? What data does each instance consist of?}
Each instance in \systemnameAPIBenchmark{} consists of a query input for an API (e.g., an image or text) along with the API's output prediction/annotation and confidence scores.
For example, one instance could be an image from the image dataset COCO~\cite{Dataset_COCO_2014}, and \{\textit{(person, 0.9), (sports ball, 0.78),
(tennis racket, 0.45)}\}, the associated annotation by Microsoft API. This means Microsoft API predicts three labels, \textit{person, sports}, and \textit{tennis racket}, with confidence scores \textit{0.9, 0.78}, and \textit{0.45}, respectively.

\textbf{How many instances are there in total (of each type, if appropriate)?}
As of 08/2022, There are a total of 1,761,417 instances in the dataset. For a breakdown by task and dataset, see Table \ref{tab:APIBenchmark:DatasetClassification} 

\textbf{Is any information missing from individual instances?} Not that the authors are aware of. 

\textbf{Are relationships between individual instances made explicit
(e.g., users’ movie ratings, social network links)? } In some of the datasets, \textit{e.g.} FER \cite{baylor2017tfx} relationships between instances are provided where applicable.

\textbf{Are there recommended data splits (e.g., training, development/validation,
testing)?} There are no recommended data splits.

\textbf{Are there any errors, sources of noise, or redundancies in the
dataset?} There are no errors or sources of noise known to the authors. 

\textbf{Is the dataset self-contained, or does it link to or otherwise rely on
external resources (e.g., websites, tweets, other datasets)?} The dataset is not self-contained and relies on a number of previously released datasets. For a comprehensive list of these datasets,

\textbf{Does the dataset contain data that might be considered confidential?} HAPI is based on existing, external datasets. It does not introduce any new data that may be considered confidential, but the authors cannot speak to the confidentiality of the external datasets. 

\textbf{Does the dataset contain data that, if viewed directly, might be offensive, insulting, threatening, or might otherwise cause anxiety?} HAPI includes predictions from MLaaS APIs. These APIs may demonstrate societal bias that could be viewed as offensive or insulting. However, the authors are not aware of any such instances in the dataset. Additionally, HAPI is based on existing, external datasets. The authors of HAPI are unaware of offensive content in these external datasets. MLaaS APIs 

\textbf{Does the dataset identify any subpopulations (e.g., by age, gender)?} The dataset does not include annotations for any subpopulations. 

\textbf{Is it possible to identify individuals (i.e., one or more natural persons), either directly or indirectly (i.e., in combination with other data) from the dataset?} HAPI is based on existing, external datasets. It does not introduce any new data that could aid in the identification of individuals, but the external datasets may include data in which it is possible to identify individuals. 

\textbf{Does the dataset contain data that might be considered sensitive in any way?} See questions above. 

\subsection{Collection Process}

\textbf{How was the data associated with each instance acquired?} For each instance  in those datasets, we have evaluated the prediction from the mainstream ML APIs from 2020 to 2022.
\systemnameAPIBenchmark{} was collected from 2020 to 2022.
For classification tasks, the predictions/annotations of each API were collected in the spring of 2020, 2021, and 2022.
For structured predictions, all APIs' predictions were collected in fall 2020 and spring 2022, separately. 
The details  can be found in Table \ref{tab:APIBenchmark:MLAPIs}.

\textbf{What mechanisms or procedures were used to collect the data
(e.g., hardware apparatuses or sensors, manual human curation,
software programs, software APIs)? } We used software APIs for MLaaS providers to collect the data.

\textbf{If the dataset is a sample from a larger set, what was the sampling
strategy (e.g., deterministic, probabilistic with specific sampling
probabilities)?} HAPI relies on a number of external datasets. It includes the full set of instances from these external datasets. The sampling strategy for each external dataset is not known to the authors of HAPI. The external datasets were chosen based on a few different criteria. First, the ML community is familiar with the datasets and they are relatively well annotated and evaluated. Second, those datasets can be easily assessed on the internet. Third, those datasets covered a diverse range of real-world scenarios (for example, the COCO dataset included objects in outdoor/indoor environments, at a small/large scale, and with different brightness). 

\textbf{Who was involved in the data collection process (e.g., students,
crowdworkers, contractors) and how were they compensated (e.g.,
how much were crowdworkers paid)?} 
The data collection process was performed by the authors of HAPI.

\textbf{Over what timeframe was the data collected?}
The MLaaS predictions were collected between 2020 and 2022. We will continue to collect predictions every six months going forward. 

\textbf{Were any ethical review processes conducted (e.g., by an institutional review board)? }
No ethical review processes were conducted. 

\textbf{Did you collect the data from the individuals in question directly,
or obtain it via third parties or other sources (e.g., websites)?}
HAPI is based on existing, external datasets which may include data collected from individuals.  

\textbf{Were the individuals in question notified about the data collection?}
The authors of HAPI are unaware of the notification policies used by the external datasets on which HAPI is based.

\textbf{Did the individuals in question consent to the collection and use
of their data?}
The authors of HAPI are unaware of the consent policies used by the external datasets on which HAPI is based.

\textbf{Has an analysis of the potential impact of the dataset and its use
on data subjects (e.g., a data protection impact analysis) been conducted?}
The authors of HAPI have not conducted any analysis of the potential impact of the dataset and its use on data subjects.

\subsection{Pre-processing/cleaning/labeling}

\textbf{Was any preprocessing/cleaning/labeling of the data done}
The preprocessing on the original inputs is as follows. On FLUENT, all 248 unique phrases were mapped to 31 unique commands as provided in the original source~\cite{Dataset_Speech_Fluent_LugoschRITB19}. 
The original labels in YELP are user ratings (1,2,3,4, and 5).
1 and 2 were  transformed to negative; 3, 4, and 5 were mapped to positive. 
IMDB, WAIMAI and SHOP contain polarized review labels and thus we directly used those labels. 
As a result, classification on all SA datasets is  a binary task.
We used a sampled version of YELP:  10,000 text paragraphs with label positive and negative separately were randomly drawn from the original YELP dataset. 
The original IMDB dataset has been partitioned into training and testing splits, and thus we used its testing split, including 25,000 text paragraphs.
All instances in WAIMAI and SHOP were used. 
The facial images in FER+ was the same as the FER dataset from the ICML 2013 Workshop on Challenges in Representation. 
A training and testing  split and regenerated labels are provided in FER+. 
We adopted the testing split with the regenerated labels.
RAFDB and AFNET contain images for both basic emotions (anger, fear, disgusting, happy, sad, surprise, and natural) and compound emotions. 
We only evaluated ML APIs on images for basic emotions, as all evaluated ML APIs focus on basic emotions. 
Different from FER+, RAFDB, and AFNET, an image in EXPW may contain multiple faces.
Thus, the labels include both the bounding box and the labelling workers' confidence. 
Thus, we extracted aligned faces as ML APIs' inputs by enlarging by 10\% and then cropping the provided face bounding boxes whose  confidence scores are larger than 0.6.

Less preprocessing was performed for structured prediction datasets. For MIC, we directly sent all raw images to the ML APIs.
A diverse collection of images is included for STR: images for advertising sales forms MTWI, while most images in ReCTS are photos taken on sing boards. LSVT's iamges are typically street view images. While all images in MTWI and ReCTS are fully annotated, LSVT contains both fully and partially annotated images. 
\systemnameAPIBenchmark{} only considers the images with full annotations as inputs to ML APIs.
For NER datasets, all samples were included in \systemnameAPIBenchmark{}.
Yet, we only focused on three widely used types of entities: person, location, and organization. 
 
Different ML APIs may use different label sets for the same tasks. 
For example, both ``disgust'' and  ``disgusting'' may be returned by different ML APIs to refer to the same  facial emotion.
Thus, label alignment is needed. For classification tasks, we manually matched each API's predicted labels to a unique number. 
For example, for FER datasets, both ``happy'' and ``happiness'' were mapped to label 3, and label 4 corresponded to ``sad'', ``sadness'', and ``unhappiness''.
For MIC with less than 100 unique labels, we were able to create the label maps manually too.  
On STR datasets, predictions (i) that are  within 0-9 or (ii) whose unicode is in the range of  u4e00-u9fff are maintained. 
For NER, we also manually mapped each API's entity type to a universal type. For example, ``people'' and ``human'' are both mapped to ``person''.

\textbf{Was the “raw” data saved in addition to the preprocessed/cleaned/labeled
data (e.g., to support unanticipated future uses)? } The raw unprocessed predictions are included and can be accessed via our Python API. 

\textbf{Is the software that was used to preprocess/clean/label the data
available?}
The software for preprocessing the data is not currently available but will be provided soon. 

\subsection{Uses}

\textbf{Has the dataset been used for any tasks already?}
\systemnameAPIBenchmark{} has been tested and used in this paper at the time of publication.
It can be used in any research related to ML prediction APIs or marketplaces, too. 
We will also maintain an incomplete list of which papers or projects have been developed on top of \systemnameAPIBenchmark{}.

\textbf{Is there a repository that links to any or all papers or systems that
use the dataset?}
As the authors become aware of papers or systems that use \systemnameAPIBenchmark{}, we will maintain a list of them on the project website  \url{https://github.com/lchen001/HAPI}.

\textbf{What (other) tasks could the dataset be used for?}
See Section~\ref{sec:conclusions} for a list of potential tasks.

\textbf{Is there anything about the composition of the dataset or the way
it was collected and preprocessed/cleaned/labeled that might impact future uses?}
The dataset relies on a limited set of existing datasets – in the future, we plan to expand the set of datasets that we use to include more diverse and up-to-date datasets. 

\textbf{Are there tasks for which the dataset should not be used?}
The authors of HAPI do not know of any particular tasks for which using this dataset should be avoided.

\subsection{Distribution}
\textbf{Will the dataset be distributed to third parties outside of the entity (e.g., company, institution, organization) on behalf of which the dataset was created?} Yes. 

\textbf{How will the dataset will be distributed (e.g., tarball on website, API, GitHub)?} The dataset is publicly available on the internet.

\textbf{When will the dataset be distributed?} The dataset is publicly available on the internet. Instructions for downloading the dataset and using the Python API are available at \url{https://github.com/lchen001/HAPI}. 

\textbf{Will the dataset be distributed under a copyright or other intellectual property (IP) license, and/or under applicable terms of use
(ToU)?} The dataset is distributed under the Apache License 2.0.

\textbf{Have any third parties imposed IP-based or other restrictions on
the data associated with the instances?}
The authors of HAPI are not aware of any third parties imposing IP-based or other restrictions on the data associated with the instances.

\textbf{Do any export controls or other regulatory restrictions apply to
the dataset or to individual instances?}
The authors of HAPI are not aware of any export controls or other regulatory restrictions applying to the dataset or to individual instances.

\subsection{Maintenance}

\textbf{Who will be supporting/hosting/maintaining the dataset?} The authors of HAPI will be supporting/hosting/maintaining the dataset.

\textbf{How can the owner/curator/manager of the dataset be contacted
(e.g., email address)?} Reach out to Lingjiao Chen (lingjiao [at] stanford [dot] edu) and Sabri Eyuboglu (eyuboglu [at] stanford [dot] edu).

\textbf{Is there an erratum?} There is currently no erratum. 

\textbf{Will the dataset be updated (e.g., to correct labeling errors, add
new instances, delete instances)?}
First, we will continuously evaluate all ML APIs considered in the paper. Currently, the evaluation is planned to occur every 6 months. If significant performance changes are consistently observed every 6 months, the update frequency will be further increased, say, to every 3 months or every month. 
MLaaS is an increasingly growing industry, and new ML APIs are launched from time to time. Thus, we plan to enlarge the set of ML APIs, datasets, and tasks in HAPI as well.  

\textbf{If the dataset relates to people, are there applicable limits on the
retention of the data associated with the instances (e.g., were the
individuals in question told that their data would be retained for
a fixed period of time and then deleted)? } HAPI is based on publicly available datasets. The retention policies of these datasets vary. 

\textbf{Will older versions of the dataset continue to be supported/hosted/maintained?} Yes.

\textbf{If others want to extend/augment/build on/contribute to the
dataset, is there a mechanism for them to do so?} Yes, potential contributors are encouraged to contact the authors of HAPI or submit a pull request on GitHub.

\end{document}